\documentclass[aps,superscriptaddress,showpacs]{revtex4}
\usepackage{graphicx,amsfonts,amsmath,color,amsbsy,amssymb,subfigure,float,appendix}
\usepackage{epstopdf}
\newcommand{\normal}{{\bf {\hat{n}}}}
\newcommand{\rPos}{{\bf r}}

\begin{document}

\title{Theoretical analysis for the optical deformation of emulsion droplets}

\author{David Tapp}
\affiliation{Department of Mathematical Sciences, Durham University, Durham, DH1 3LE, UK.}

\author{Jonathan M. Taylor}
\affiliation{School of Physics and Astronomy, University of Glasgow, Glasgow, G12 8QQ, UK.}
\affiliation{Department of Physics, Durham University, Durham, DH1 3LE, UK.}

\author{Alex S. Lubanksy}
\affiliation{Department of Engineering Science, University of Oxford, Oxford, OX1 3PJ, UK.}
\affiliation{School of Engineering, Edith Cowan University, Joondalup, WA 6027, Australia.}

\author{Colin D. Bain}
\affiliation{Department of Chemistry, Durham University, Durham, DH1 3LE, UK.}

\author{Buddhapriya Chakrabarti}
\email{buddhapriya.chakrabarti@durham.ac.uk}
\affiliation{Department of Mathematical Sciences, Durham University, Durham, DH1 3LE, UK.}
\affiliation{The Isaac Newton Institute of Mathematics, Cambridge University, Cambridge, CB3 OEH, UK.}


\begin{abstract}
We propose a theoretical framework to predict the three-dimensional shapes of optically deformed micron-sized emulsion droplets with ultra-low interfacial tension. The resulting shape and size of the droplet arises out of a balance between the interfacial tension and optical forces. Using an approximation of the laser field as a Gaussian beam, working within the Rayleigh-Gans regime and assuming isotropic surface energy at the oil-water interface, we numerically solve the resulting shape equations to elucidate the three-dimensional droplet geometry. We obtain a plethora of shapes as a function of the number of optical tweezers, their laser powers and positions, surface tension, initial droplet size and geometry. Experimentally, two-dimensional droplet silhouettes have been imaged from above, but their full side-on view has not been observed and reported for current optical configurations. This experimental limitation points to ambiguity in differentiating between droplets having the same two-dimensional 
projection but with disparate three-dimensional shapes. Our model elucidates and quantifies this difference for the first time. We also provide a dimensionless number that indicates the shape transformation (ellipsoidal to dumbbell) at a value $\approx 1.0$, obtained by balancing interfacial tension and laser forces, substantiated using a data collapse.
\end{abstract}

\pacs{Nonlinear Optics at Surfaces, Optics at surfaces, Surfaces, Optical tweezers or optical manipulation.}
\maketitle
\section{Introduction}
Optical manipulation of fluid interfaces having ultra-low interfacial tension was realised recently in systems of micron-sized emulsion droplets~\cite{Bain:06,Bain:11}. Potential applications of such techniques lie in pumping fluids, performing reaction chemistry at the attolitre scale and understanding the behaviour of oil droplets in surfactant-enhanced oil recovery~\cite{Karlsson:06, Hirasaki:11}. Similar techniques have been applied to deform cells, which have a non-zero bending energy of the lipid bilayer in addition to the usual interfacial tension~\cite{Bronkhorst:95, Guck2000a, Guck2001, Dharmadhikari:04}.

The precision and non-destructive nature of manipulating particles by optical tweezers can be a benefit over standard mechanical techniques. In most physically realisable situations the radiation pressure exerted by the beam on the trapped particle is orders of magnitude weaker than the Young's modulus of the solid or the Laplace pressure of the confined fluid. As a result, particles do not deform in the trap: a solid particle retains its initial shape while a fluid droplet assumes a spherical geometry. Therefore, historically many applications of optical tweezing have been limited to exerting~\cite{Ashkin:86,Neuman2004,Hansen:05,McGloin2006} and measuring~\cite{Block1990,gibson:accuracy,Davenport2000} external forces on a rigid trapped object. The situation is, however, different for fluids with low interfacial tension, where the object itself can be deformed in response to the laser field.

Ashkin and Dziedzic reported small but measurable deformations on planar soft surfaces using a focused and pulsed laser beam~\cite{Ashkin:73}. Further, Casner and Delville showed that large deformations could be observed if the interfacial tension was lowered~\cite{Casner:01}. It is known that the interfacial tension can be lowered substantially by the addition of surfactants. Ward \textit{et al.}\ showed that the interfacial tension of heptane droplets could be lowered to $\gamma \approx 10^{-5} - 10^{-6}\,N m^{-1}$ by the addition of suitably selected surfactants~\cite{Bain:06}. As a result, they were able to deform the emulsion droplets into various shapes using multiple continuous-wave lasers. Our theoretical work presented here is motivated by these experiments.

The problem is similar in spirit to that of lipid vesicles deformed via micro-manipulation techniques~\cite{Evans:89} where a bending free energy of the lipid bilayer is balanced against externally applied forces to give its resulting shape. The difference between the membrane and fluid geometry arises from the nature of the constraints imposed on the shape equations. In the vesicle case, the total area is conserved while for a fluid droplet the volume of enclosed fluid is constant throughout the deformation process. Furthermore, based on dimensional arguments one can show that so long as $\frac{\kappa}{\xi^{2}} < \gamma$, where $\kappa$ is the bending modulus and $\xi$ is a length scale based on the radius of curvature of the droplet surface, in the hydrodynamic limit the interfacial tension contribution dominates over the curvature energy contribution.

We propose a numerical model that predicts the three-dimensional steady-state shapes of droplets with ultra-low interfacial tension under the influence of one or more optical traps. Several authors have recently published models exploring different regimes of optical deformation of liquid droplets~\cite{Moller:09,Ellingsen:13}. The key feature of the model we propose here is that it does not assume small, linear deformation of the droplet. Our model makes no assumptions on the final shape of the droplet and there is no restriction that the optical traps have to be focused at the centre of the droplet. Our model allows us to investigate the equilibrium shapes of droplets as a function of different parameters, such as laser power $\left(P_{0}\right)$, numerical aperture $\left(NA\right)$, initial droplet radius $\left(R_{d}\right)$, interfacial tension $\left(\gamma\right)$ and the parameter $\left(\tilde{n}\right)$ which is related to the ratio of refractive indices as $\tilde{n} = 1 - \frac{n_{2}}{n_{1}}$, 
where $n_{1}$ and $n_{2}$ are the refractive indices of the droplet and external media, respectively. From these results we have defined a dimensionless deformation number $N_{d}$ which is a function of $P_{0}$, $NA$, $R_{d}$, $\gamma$ and $\tilde{n}$. Using this dimensionless number we are able to predict previously undescribed shape transformations of a droplet in a single optical trap.

The mathematical model is described in the Section~\ref{maths}. This is followed by an outline of the numerical implementation of the model in Section~\ref{numerics}. We then present and discuss the results of the deformation of a droplet in single and multiple optical traps. Finally, we discuss future applications and extensions to the present model.

\section{Mathematical Model for Droplet Deformation}
\label{maths}
In our model we describe the surface of the droplet (the interface between the liquid droplet and its host medium) in terms of a single-valued function $R(\theta,\phi)$ in spherical polar coordinates, where $R$ defines the distance between the interface and a preassigned fixed origin. In the absence of any external forces, a liquid droplet assumes a spherical shape, as  a result of two antagonistic forces: the interfacial tension which tends to minimise the area, and the bulk pressure of the internal fluid which translates into a force acting along the local normal to an infinitesimal surface element $dS$. The energy function for an isolated droplet can be written as:
\begin{equation}\label{isolated_droplet}
E = \oint_{A} \gamma dS - \int_{V} P_{int} dV
\end{equation}
Thus, the internal pressure for an isolated droplet at equilibrium turns out to be $P_{int} = \frac{2\gamma}{R_{d}}$. In the presence of one or more optical tweezers, the total pressure acting at a position $\rPos{}$ on an infinitesimal surface element $dS$ at the oil-water interface can be written as:
\begin{equation}\label{pressure-eq}
P(\rPos{}) = P_{lap}(\rPos{}) + P_{opt}(\rPos{}) + P_{int}
\end{equation}
where $P_{lap}$ is the Laplace pressure, $P_{opt}$ is the optical pressure due to momentum transfer from the laser field to the interface, and $P_{int}$ is the internal pressure within the droplet, which we will treat as an unknown variable dependent on the specific experimental configuration, but uniform throughout the droplet since we are only considering equilibrium structures.

The Laplace pressure $P_{lap}(\rPos{})$ acting on a surface element on the oil-water interface (interfacial tension $\gamma$), at a position $\rPos{}$ and with outward normal $\normal{}(\rPos{})$, is calculated using the Young-Laplace equation:
\begin{equation}
\label{laplace-pressure-equation} P_{lap}(\rPos{}) = \gamma \nabla \cdot \normal{}(\rPos{})
\end{equation}
from using Gauss' theorem on the first integral in Eq.~\ref{isolated_droplet}. Note that by convention this pressure acts inwards in the case of a convex droplet surface.

The optical pressure $P_{opt}(\rPos{})$ across the oil-water interface is calculated by considering the momentum transferred to the interface as light is reflected and refracted at the interface. An exact solution for the light field in the presence of a dielectric object could be calculated in the form of a series expansion within the framework of T-matrix theory~\cite{nonspherical-review}. However, due to the excessive computational demands that this would impose, we instead adopt a localized ray-optics approximation similar to that described in~\cite{Xu2009}. We generalize the treatment to an arbitrary shape, but at the same time we note that the refractive index ratio $m=\frac{n_{2}}{n_{1}}$ between the oil and water phases is close to one, such that the phase shift of rays crossing the droplet is small compared to the wavelength of the trapping laser and we are close to the Rayleigh-Gans regime~\cite{vanDeHulst}. We will therefore make the approximation that the field is unperturbed by the presence of 
the droplet and calculate the localized Fresnel reflection and refraction for each surface element using the unperturbed laser field, without considering higher-order reflections.

If we consider a surface element on the oil-water interface, at a position $\rPos{}$ and with outward normal $\normal{}(\rPos{})$, then for an incident beam with momentum density $p_0 n_1{\bf {\hat{s}}}$, where ${\bf {\hat{s}}}$ is the Poynting vector, we can apply standard laws of reflection and refraction~\cite{Herzberger:58} and, after some algebraic manipulation, we find the optical pressure acting on the surface (in the direction of the outward normal) to be:
\begin{equation}\label{optical-pressure-eq}
P_{opt}(\rPos{}) = -p_0 n_1 \left((2F_r + F_t) \mu - F_t  {\rm sgn}(\mu)\sqrt{(n_2/n_1)^2-1+\mu^2} \right)
\end{equation}
where $\mu = {\bf {\hat{s}}} \cdot {\bf {\hat{n}}}$, $F_r$ and $F_t$ are the Fresnel power reflection and transmission coefficients for the angle of incidence $\theta_{inc}=\arccos (|\mu|)$, and $n_1$ and $n_2$ are the refractive indices for the media in which the incoming and refracted beams are propagating.

We also require a description for the momentum density $p_0 n_1{\bf {\hat{s}}}$ of the beam. The fact that our droplets are several wavelengths in diameter, significantly larger than the trapping beam waist, implies that a series expansion around the beam focus, such as that presented in~\cite{barton:5th-order} and used in surface optical stress calculations in~\cite{Xu2009}, is not suitable here. However, observing that for the droplet radii we are interested in the interface is generally several Rayleigh lengths from the (tight) laser focus, it becomes apparent that a far-field scalar model of a tightly focused Gaussian beam is appropriate~\cite{Siegman1986}.

\begin{figure}[htbp]
\centering
\subfigure[][]{\includegraphics[width=0.40\textwidth , trim = {2cm 3cm 2cm 1cm},clip = true]{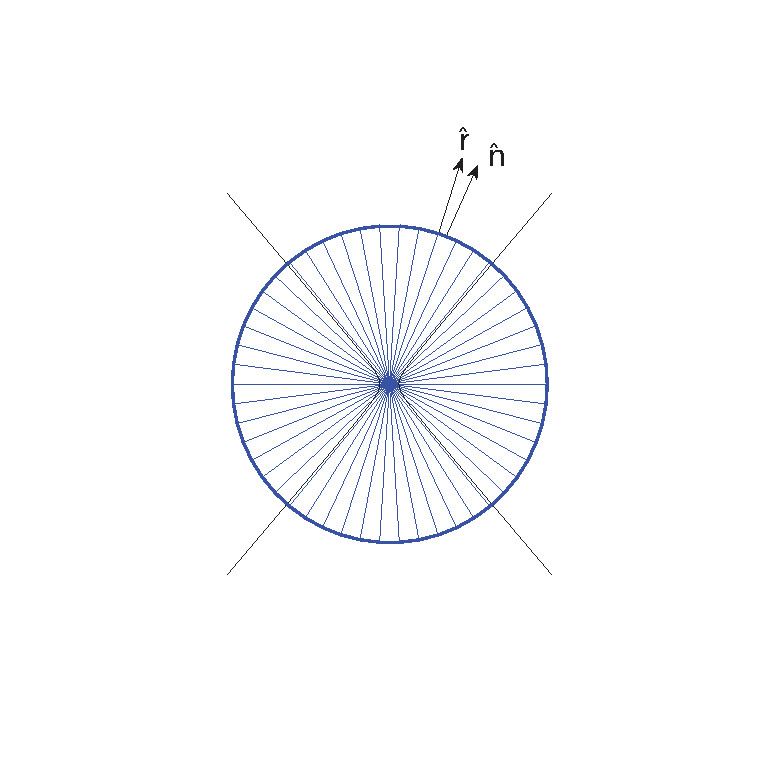}}\centering
\subfigure[][]{\includegraphics[width=0.40\textwidth , trim = {2cm 3cm 2cm 1cm},clip = true]{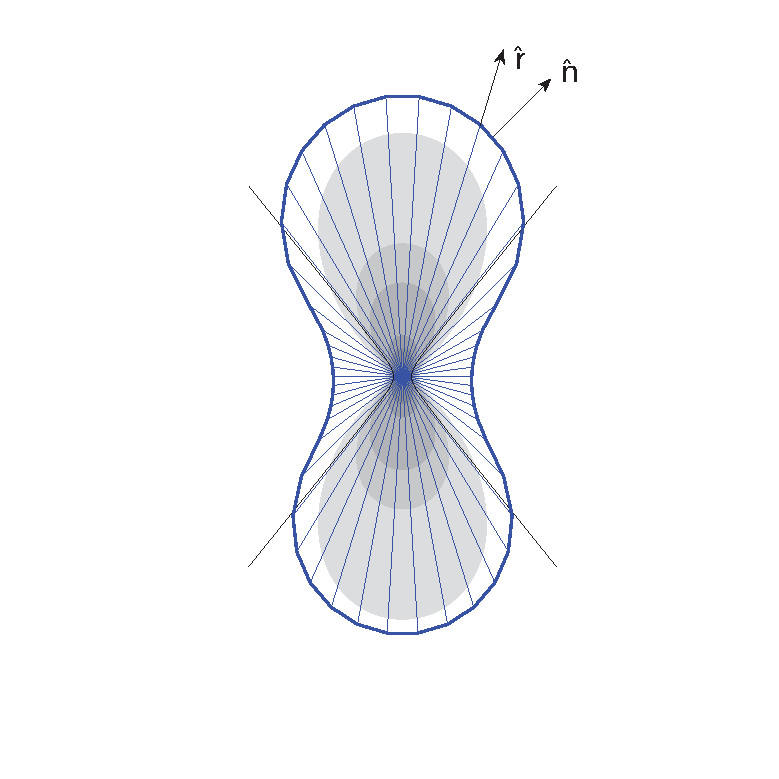}}\centering
\caption{Droplet configurations corresponding to the ``spoke'' model for (a) an undeformed droplet of $R_{d} = 5.0\,\mu m$ with $\gamma = 10^{-6}\,Nm^{-1}$ and (b) the corresponding converged shape in an optical trap with $P_{0} = 0.20\,W$ and numerical aperture $NA = 1.20$. The Gaussian beam is represented by the black lines, and part of the intensity distribution is also shown. The radial direction $(\hat{{\bf r}})$ of a ``spoke'' is also shown, along with the corresponding interface normal $(\hat{{\bf n}})$.}
\label{fig:modelfig1}
\end{figure}

We can now return to the pressure acting on the surface of the droplet (Eq.~\ref{pressure-eq}). In order to obtain equilibrium droplet shapes we employ a relaxation dynamics scheme at each ``spoke'' of the coordinate system. Thus the time evolution of each point on the surface of the droplet is given by:
\begin{equation}\label{equation-of-motion}
\frac{dR(\theta,\phi)}{dt} = \beta( P_{opt}(\theta,\phi)  - P_{lap}(\theta,\phi) + P_{int}) / (\hat{{\bf n}} \cdot \hat{{\bf r}})
\end{equation}
where the $(\hat{{\bf n}} \cdot \hat{{\bf r}})$ term takes account of the fact that the interface normal is not collinear to the (fixed) radial direction along which $R(\theta,\phi)$ is measured -- see Figure \ref{fig:modelfig1}, and hence $R$ will in general vary faster than the rate of normal motion of the interface. The friction coefficient $\beta$ dictates the inverse time scale for a droplet to relax to equilibrium. This, however, cannot be chosen to be arbitrarily large as it compromises the stability of the numerical scheme used to integrate Eq.~\ref{equation-of-motion}. This equation of motion represents overdamped motion. Thus our model does not attempt to predict the detailed \emph{transient} motion of the droplet surface over time as it approaches its final shape, since we have deliberately selected a simplified equation of motion that does not take complete account of the hydrodynamic environment of the droplet. Although a final solution to our model represents a physically correct \emph{steady-
state solution} for a given set of experimental parameters, we note that the ``time'' variable in our calculations does not represent real-world time, the trajectory taken through parameter space to attain the final solution is not physically rigorous, and if experimental parameters support multiple steady-state solutions then our model may not be able to predict for certain which solution represents the kinetic product from a given set of initial conditions.

The equation of motion is integrated with respect to time in order to determine the physically correct steady-state solution, subject to a constant-volume constraint. The volume of the droplet is defined as:
\begin{equation}\label{volume-integral}
V = \frac{1}{3} \int_S R(\theta,\phi)^3 \sin\theta d\theta d\phi
\end{equation}
and hence the boundary condition can be expressed as the constraint:
\begin{equation}\label{volume-integral-constraint}
\frac{dV}{dt} = 0 = \int_S R(\theta,\phi)^2 \frac{dR(\theta,\phi)}{dt} \sin\theta d\theta d\phi
\end{equation}
Substituting Eq.~\ref{equation-of-motion} into this, we obtain an expression for the free parameter $P_{int}$ representing the internal pressure of the droplet:
\begin{equation}\label{internal-pressure-equation}
P_{int} = \left(\int_S \frac{R(\theta,\phi)^2 (P_{lap} + P_{opt})}{\hat{{\bf n}} \cdot \hat{{\bf r}}} \sin\theta d\theta d\phi\right) / \left( \int_S \frac{R(\theta,\phi)^2}{\hat{{\bf n}} \cdot \hat{{\bf r}}} \sin\theta d\theta d\phi \right)
\end{equation}

For completeness we note that once the shape of the droplet has converged, as a result of these pressure contributions, by removing the presence of the optical tweezers from our model we return to the energy function of an isolated droplet (Eq.~\ref{isolated_droplet}). As a result the deformed droplet converges back to a perfect sphere with radius $R_{d}$.

\section{Numerical Implementation}
\label{numerics}
In our model the continuous surface of the droplet is sampled at a certain number of discrete points. These points lie on a regular grid in spherical coordinates, with $N_\theta \times N_\phi$ nodes equally spaced in $(\theta, \phi)$ space. Node (i,j) lies at coordinates $(\theta_i=\frac{i+0.5}{N_\theta}\pi, \phi_j=\frac{j+0.5}{N_\phi}2\pi)$ for $0 \leq i < N_\theta$ and $0 \leq j < N_\phi$.  For each node a radius $R_{ij}$ is defined, each representing a point $(R_{ij}, \theta_i, \phi_j)$ on the surface of the droplet (the oil-water interface), so that the radius of each node defines a ``spoke'' extending from the origin in the direction $(\theta,\phi)$ -- see Figure \ref{fig:modelfig1}. We then solve the equation of motion (Eq.~\ref{equation-of-motion}) on this discretized grid, using finite-difference expressions as an estimate for local derivatives on the surface of the droplet.

We can calculate the surface normal  $\hat{\mathbf{n}}$ as follows~\cite{Spivak:75}:
\begin{equation}\label{surface-normal-equation}
\hat{\mathbf{n}}=\frac{\nabla\left(r-R\left(\theta,\phi\right)\right)}{\left|\nabla\left(r-R\left(\theta,\phi\right)\right)\right|}
\end{equation}
where $r$ is the radial coordinate and $R\left(\theta,\phi\right)$ is the radius of a ``spoke'' at a given $(\theta$,$\phi)$ coordinate.

At first glance it might appear convenient to implement Eq.~\ref{laplace-pressure-equation} and \ref{surface-normal-equation} in a two-stage finite-difference scheme, however in practice this two-stage approach leads to a decoupling between odd and even points in the grid, which is difficult to eliminate. The solution is to derive a single expression for the Laplace pressure directly in terms of derivatives of the radius. This combined expression proves to be rather complicated in spherical coordinates:
\begin{eqnarray}\label{laplace-eq-with-derivatives}
P_{lap}(R, \theta, \phi)&\hspace{-2mm}=\hspace{-2mm}& \gamma \left[ \frac{2u}{R} - \frac{u}{R^2} \left( \cot\theta \frac{\partial R}{\partial \theta} + \frac{\partial^2 R}{\partial \theta^2} + \csc^2\theta \frac{\partial^2 R}{\partial \phi^2} \right) \right.  \\
&& \hspace{5mm} + \frac{u^3}{R^3} \left( \left( 1 +  \frac{1}{R} \frac{\partial^2 R}{\partial \theta^2} \right)\left(\frac{\partial R}{\partial\theta}\right)^2 \right. \nonumber \\
&& \hspace{15mm} + \left(  \csc^2\theta  - \frac{\cos\theta}{R \sin^3\theta} \frac{\partial R}{\partial \theta} \right) \left(\frac{\partial R}{\partial\phi}\right)^2 \nonumber \\
&& \hspace{15mm} \left. \left. + \frac{2}{R \sin^2\theta} \frac{\partial R}{\partial\theta}\frac{\partial R}{\partial\phi}\frac{\partial^2 R}{\partial\theta\partial\phi} + \frac{1}{R\sin^4\theta} \left(\frac{\partial R}{\partial\phi}\right)^2 \frac{\partial^2 R}{\partial\phi^2} \right) \right] \nonumber,
\end{eqnarray}
where:
\begin{eqnarray}
u &\hspace{-2mm}=\hspace{-2mm}& \left( 1 + \frac{1}{R^2}\left(\frac{\partial R}{\partial\theta}\right)^2 + \frac{1}{R^2\sin^2\theta} \left(\frac{\partial R}{\partial\phi}\right)^2 \right)^{-\frac{1}{2}}.
\end{eqnarray}
Now, by evaluating Eq.~\ref{optical-pressure-eq}~and~\ref{laplace-eq-with-derivatives} with the help of second-order-accurate finite-difference expressions, and numerically integrating Eq.~\ref{internal-pressure-equation}, we are able to numerically integrate the equation of motion (Eq.~\ref{equation-of-motion}) with respect to time, using the Runge-Kutta-Fehlberg method, until motion ceases and the droplet has converged to a stable shape.

\begin{figure}[htbp]
\centering
\subfigure[][]{\includegraphics[width=0.3\textwidth, trim=1cm 0cm 1.5cm 0.5cm,clip=true]{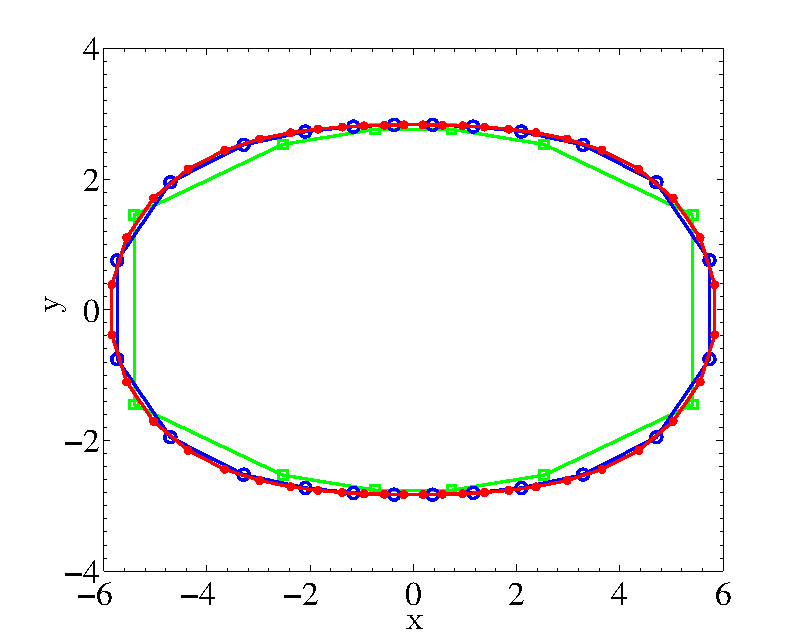}}\centering
\subfigure[][]{\includegraphics[width=0.3\textwidth, trim=1cm 0cm 1.5cm 0.5cm,clip=true]{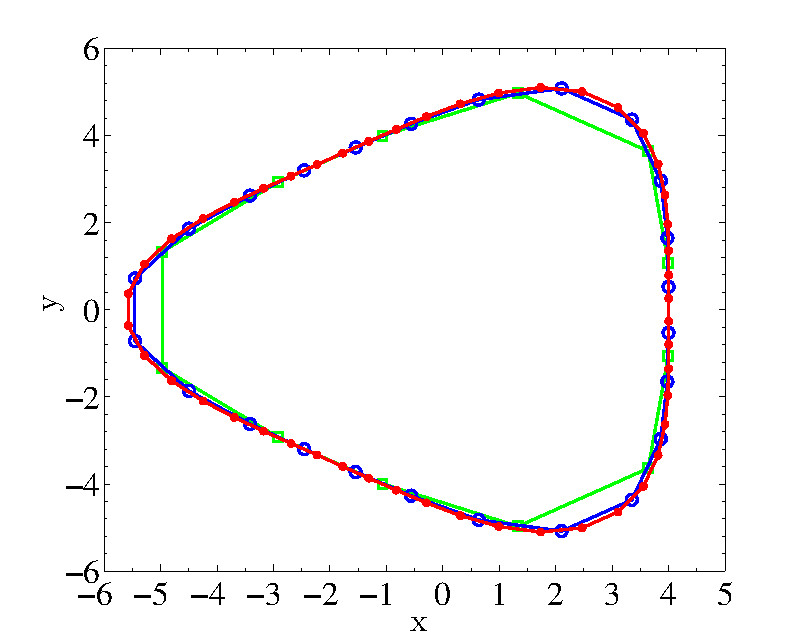}}\centering
\subfigure[][]{\includegraphics[width=0.3\textwidth, trim=1cm 0cm 1.5cm 0.5cm,clip=true]{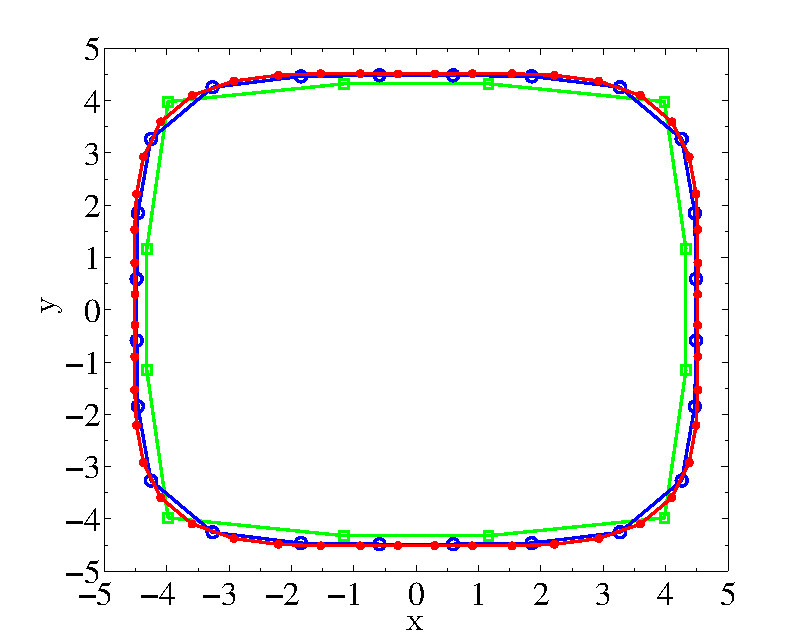}}\centering
\caption{Comparison between a very coarse mesh of $N_{\theta}=13$ \& $N_{\phi}=12$ points (green), a coarse mesh of $N_{\theta}=25$ \& $N_{\phi}=24$ points (blue) and a refined mesh of $N_{\theta}=51$ \& $N_{\phi}=48$ (red), for a $R_{d} = 5.0\, \mu m$ droplet deformed in (a) two optical traps, (b) three optical traps and (c) four optical traps.}
\label{fig:coarsevsfine}
\end{figure}

To improve computational efficiency we initially use a mesh size of $N_{\theta}=25$ and $N_{\phi}=24$ nodes. Once the droplet shape has converged using this coarse mesh, we refine to $N_{\theta}=51$ and $N_{\phi}=48$. Figure \ref{fig:coarsevsfine} shows a comparison between the converged shapes of a droplet with $R_{d} = 5.0\,\mu m$ in an increasing number of optical traps. It can be seen that the refinement improves the shapes of the droplets in the regions of high curvature where the optical traps are positioned, whilst only minor improvements can be observed for the rest of the droplet. A mesh size of $N_{\theta}=13$ and $N_{\phi}=12$ is also shown in green in Figure \ref{fig:coarsevsfine}. Such a coarse mesh is incapable of ensuring a smooth surface is obtained, while the initial mesh size of $N_{\theta}=25$ and $N_{\phi}=24$ nodes represents a good compromise between computational demands and accuracy, being able to give a reasonable initial estimate of the droplet shape that can later be refined. 
Furthermore, the calculation of the Laplace pressure would be erroneous with an insufficient number of mesh points. Our choice of $N_{\theta}$ and $N_{\phi}$ is also verified by considering the well-known fact that the integral of the mean curvature $H$ over a closed surface $S$ is equal to zero~\cite{Blackmore:85}:
\begin{equation}
\int \!\!\! \int_{S} H \cdot \hat{{\bf n}} dA= 0
\end{equation}
As the number of $N_{\theta}$ and $N_{\phi}$ nodes is increased in our model, this calculated quantity will tend to zero.

\section{Results and Discussion}
\label{results}
Continuous-wave, unpolarized, lasers having a wavelength $\lambda=1064\,nm$ have been modelled for results reported in this paper. Measured values of the refractive index of a heptane droplet and an external water medium, $n_{1}=1.38$ and $n_{2}=1.33$ respectively, have been used in our calculations, unless otherwise stated.

\subsection{Single Optical Trap}
We begin our discussion with the deformation of a droplet in a single optical trap. The optical trap is positioned such that the centre of the droplet initially coincides with the focal point of the beam. The steady-state profiles are obtained when the ``spoke'' positions parameterising the surface do not change as a function of time.

\begin{figure}[htbp]
\centering
\subfigure[][$P_{0}=0.02\,W$]{\includegraphics[width=0.3\textwidth, trim=0cm 0cm 0cm 0cm,clip=true]{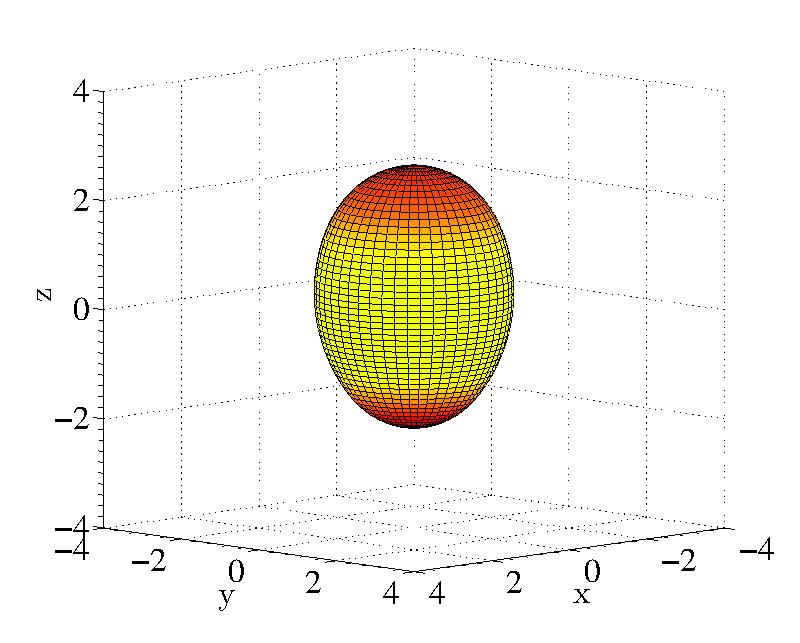}}\centering
\subfigure[][$P_{0}=0.04\,W$]{\includegraphics[width=0.3\textwidth, trim=0cm 0cm 0cm 0cm,clip=true]{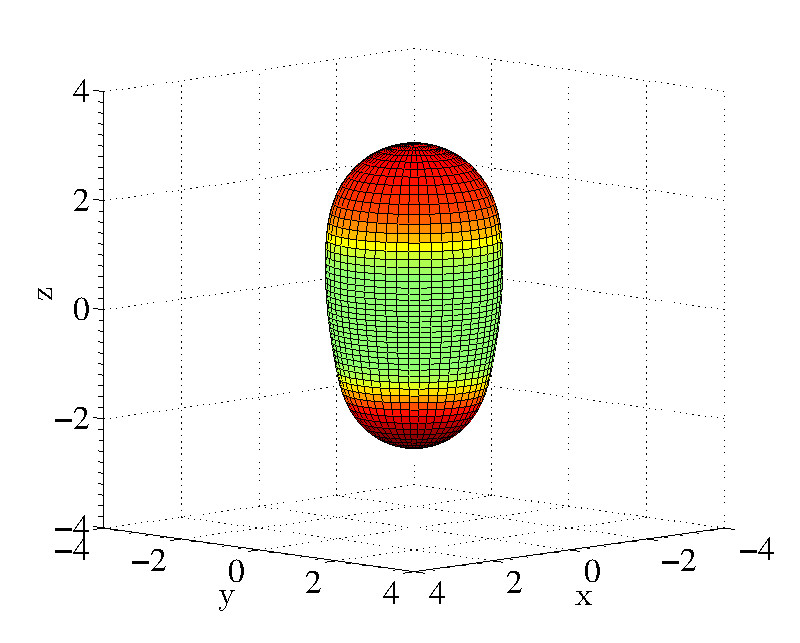}}\centering
\subfigure[][$P_{0}=0.06\,W$]{\includegraphics[width=0.3\textwidth, trim=0cm 0cm 0cm 0cm,clip=true]{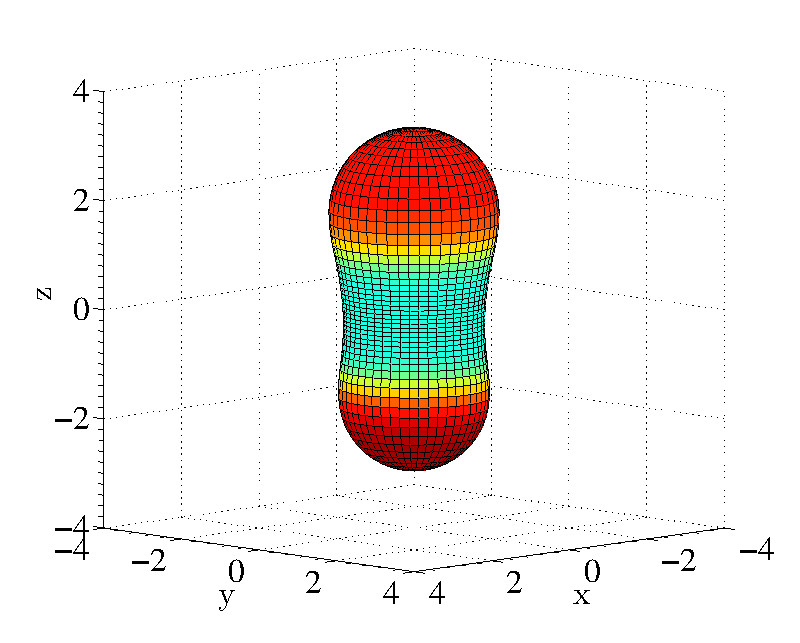}}\centering

\subfigure[][$P_{0}=0.08\,W$]{\includegraphics[width=0.3\textwidth, trim=0cm 0cm 0cm 0cm,clip=true]{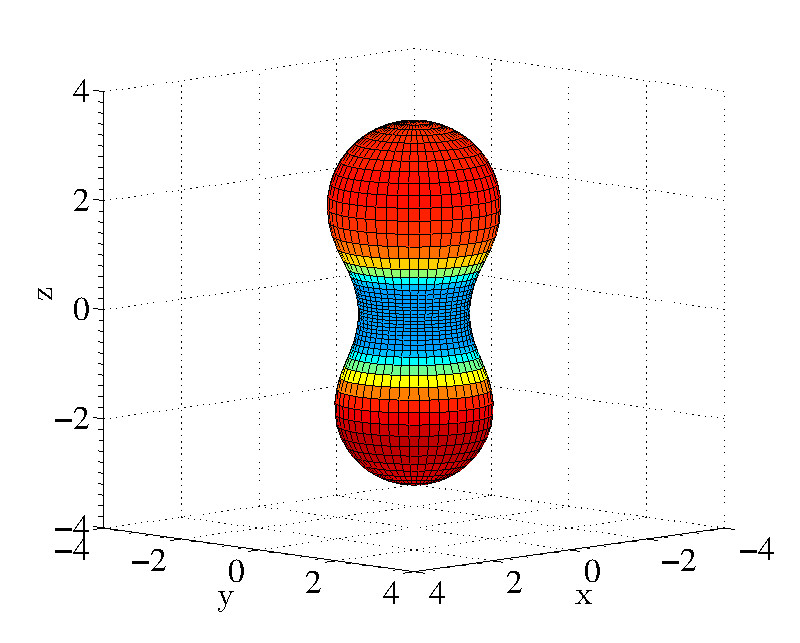}}\centering
\subfigure[][$P_{0}=0.10\,W$]{\includegraphics[width=0.3\textwidth, trim=0cm 0cm 0cm 0cm,clip=true]{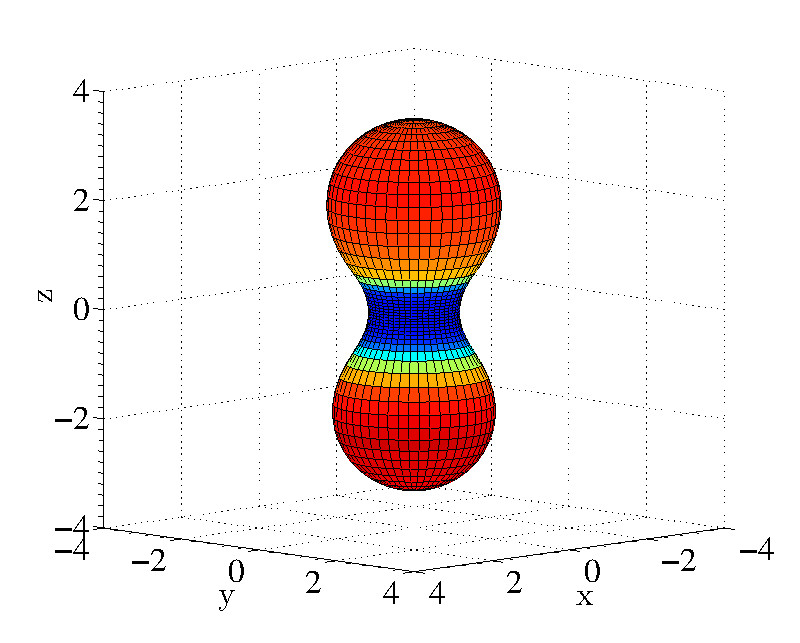}}\centering
\subfigure[][$P_{0}=0.12\,W$]{\includegraphics[width=0.3\textwidth, trim=0cm 0cm 0cm 0cm,clip=true]{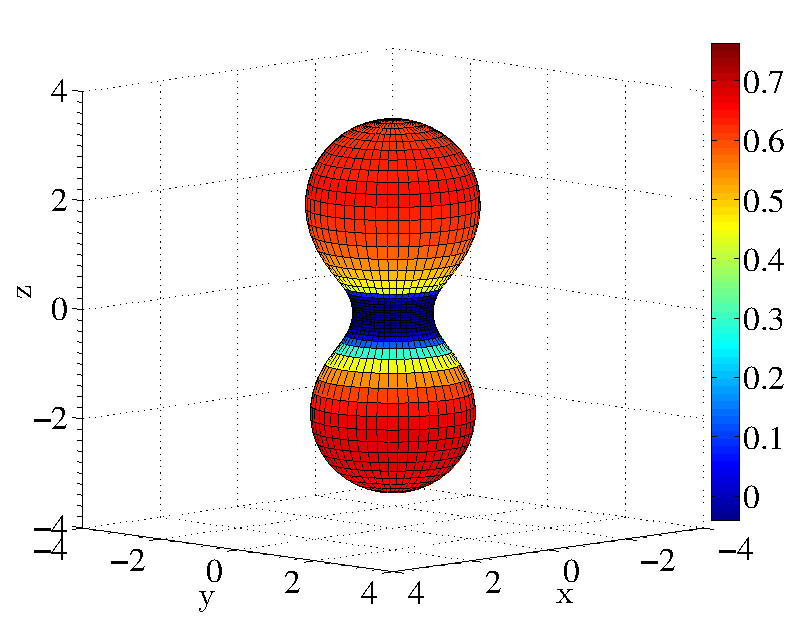}}\centering
\caption{Deformed conformations of an initial spherical droplet with radius $R_{d}=2.0\,\mu m$ and interfacial tension $\gamma = 10^{-6}\,Nm^{-1}$ in a single optical trap as a function of increasing laser power $P_{0}$. The colour scheme represents the variation in mean curvature, in units of $\mu m^{-1}$.}
\label{fig:1laser_results1}
\end{figure}

Figure \ref{fig:1laser_results1} shows the converged three-dimensional shape of a deformed droplet with initial spherical geometry $R_{d}=2.0\,\mu m$, and a surface tension $\gamma = 10^{-6}\,Nm^{-1}$, as a function of increasing laser power $P_{0}$. The radius of the beam waist for each optical trap was kept constant at $w_{0}=0.282\,\mu m$, which corresponds to a numerical aperture of $NA=1.20$. As the laser power increases the droplet elongates to assume a lozenge form, with its long axis parallel to the direction of propagation of light (taken to be along the $+z$ axis). For laser powers $P_{0} \gtrsim 0.055\,W$ the droplet has a dumbbell-like shape with an hour-glass connecting two spherical caps. For these configurations one of the principal curvatures is negative. Experimental observations of these deformations have so far been limited to their two-dimensional projections along the axis of laser propagation~\cite{Bain:06}. Note that the shapes are asymmetric with respect to $z$. If we consider the 
gradient force alone then we would expect a symmetric shape, reflecting the symmetry of the intensity distribution in the trapping laser field. However the scattering force (which acts in the local direction of propagation of the laser field) breaks this symmetry and pushes the interface, and hence the droplet, in the direction of propagation of the laser and leads to ``bulging'' of the droplet beyond the laser focus.

The colour scheme used in Figure \ref{fig:1laser_results1} indicates the variation in the calculated mean curvature over the surface of the droplet. For low laser powers the mean curvature is close to $0.5\,\ \mu m^{-1}$ corresponding to that of a perfect sphere with radius $R_{d} = 2.0\,\mu m$, except at the ``tips'' of the droplet. As the laser power is increased we see a larger variation in the mean curvature, with the lowest mean curvature at the waist of the droplet, and the highest mean curvature at the two spherical caps. For laser powers greater than $P_{0} = 0.12\,W$ the mean curvature at the waist becomes negative.

\begin{figure}[htbp]
\centering
\subfigure[][]{\includegraphics[width=0.45\textwidth, trim=0cm 0cm 1cm 0cm,clip=true]{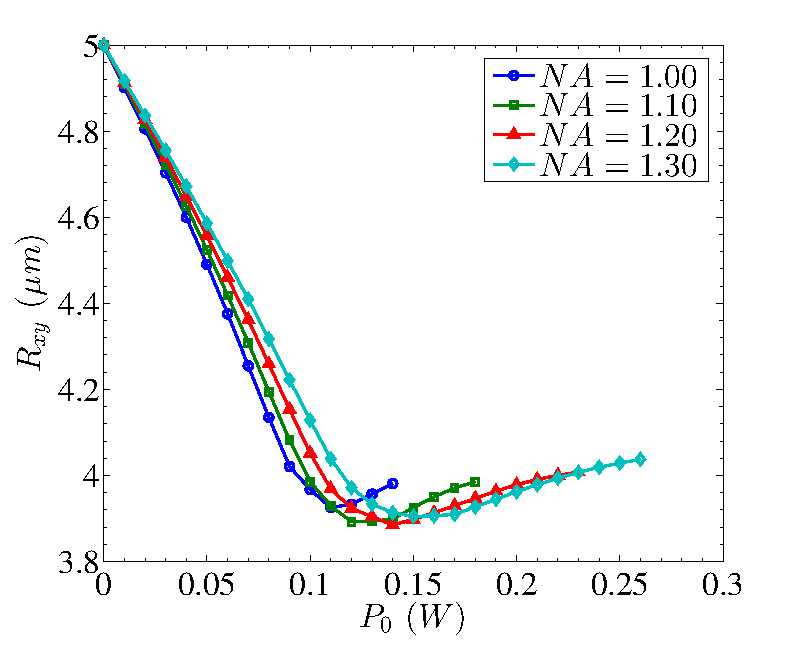}}\centering
\subfigure[][]{\includegraphics[width=0.45\textwidth, trim=0cm 0cm 1cm 0cm,clip=true]{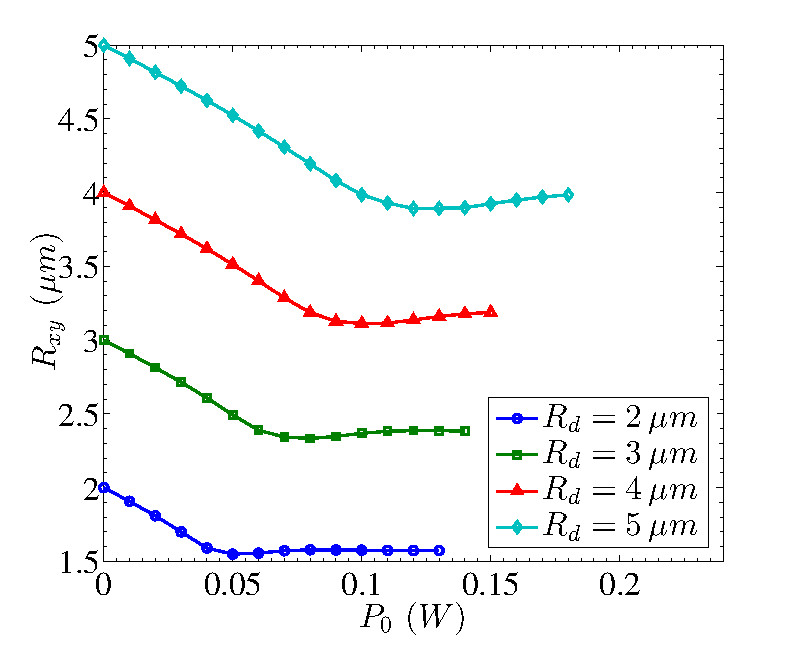}}\centering

\subfigure[][]{\includegraphics[width=0.45\textwidth, trim=0cm 0cm 1cm 0cm,clip=true]{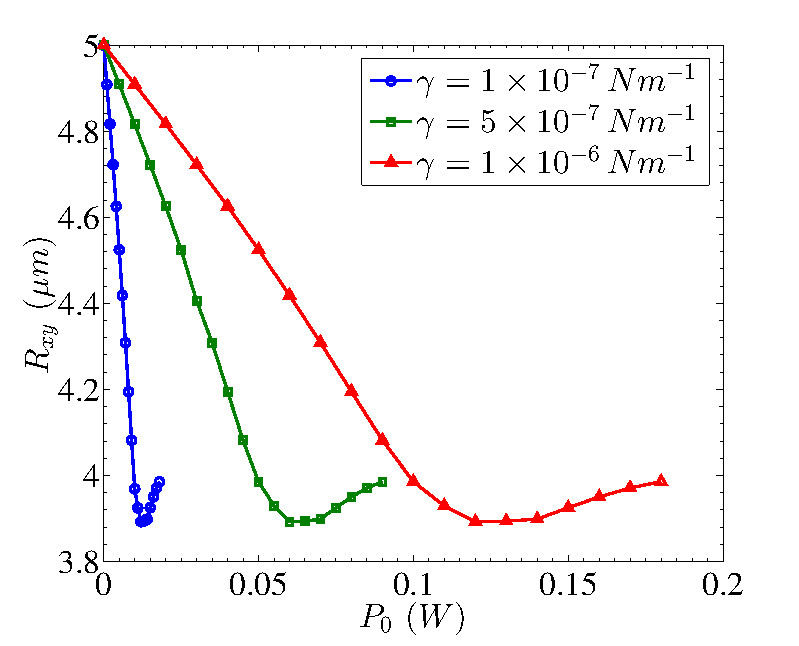}}\centering
\subfigure[][]{\includegraphics[width=0.45\textwidth, trim=0cm 0cm 1cm 0cm,clip=true]{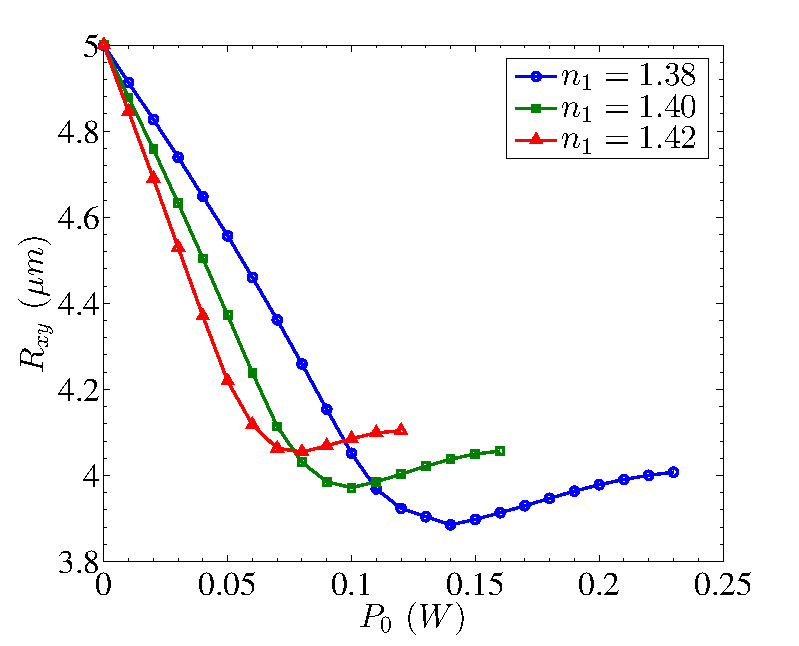}}
\caption{Variation of the maximum projected radius of a droplet as a function of (a) $NA$ for $R_{d}=5.0\,\mu m$ and $\gamma=10^{-6}\,Nm^{-1}$, (b) initial droplet radius $R_{d}$ with $NA=1.10$ and $\gamma = 10^{-6}\,Nm^{-1}$, (c) interfacial tension $\gamma$ for $R_{d}=5.0\,\mu m$ with $NA = 1.10$ and (d) refractive index $n_{1}$ of the droplet with $R_{d}=5.0\,\mu m$ and $\gamma=10^{-6}\,Nm^{-1}$ and an optical trap with $NA = 1.20$.}
\label{fig:onelasergraphs}
\end{figure}

Since current experimental observations are limited to two-dimensional $xy$ projections, the deformation of these droplets in a single optical trap is only known due to a decrease in the maximum projected radius~\cite{Bain:06}. The variation of the maximum projected droplet radius $R_{xy}$ as a function of the laser power $P_{0}$ for a constant value of interfacial tension $\gamma=10^{-6}\,Nm^{-1}$ and initial droplet size $R_{d}=5.0\,\mu m$ for different values of numerical aperture $NA$ is shown in Figure \ref{fig:onelasergraphs}a. $R_{xy}$ changes non-monotonically as a function of $P_{0}$, with the minimum occurring at the transition to a dumbbell-like form, a transition that we will discuss in more detail below. The initial linear decrease in the slope of the $R_{xy}$ vs.\ $P_{0}$ curve is due to the elongation along the $z$-axis. Beyond the transition point, the equilibrated configurations show a narrowing of the neck region connecting the two spherical caps. A similar trend is observed for different
values of the initial droplet size $R_{d}$ as shown in Figure \ref{fig:onelasergraphs}b. Figures \ref{fig:onelasergraphs}c \& d show the similar qualitative trends for a droplet with $R_{d} = 5.0\,\mu m$ for different values of the interfacial tension $10^{-7} \lesssim \gamma \lesssim 10^{-6}\,Nm^{-1}$ and refractive index of a droplet $n_{1}$ respectively.

\begin{figure}[htbp]
\centering
\subfigure[][]{\includegraphics[width=0.45\textwidth, trim=0cm 0cm 1cm 0cm,clip=true]{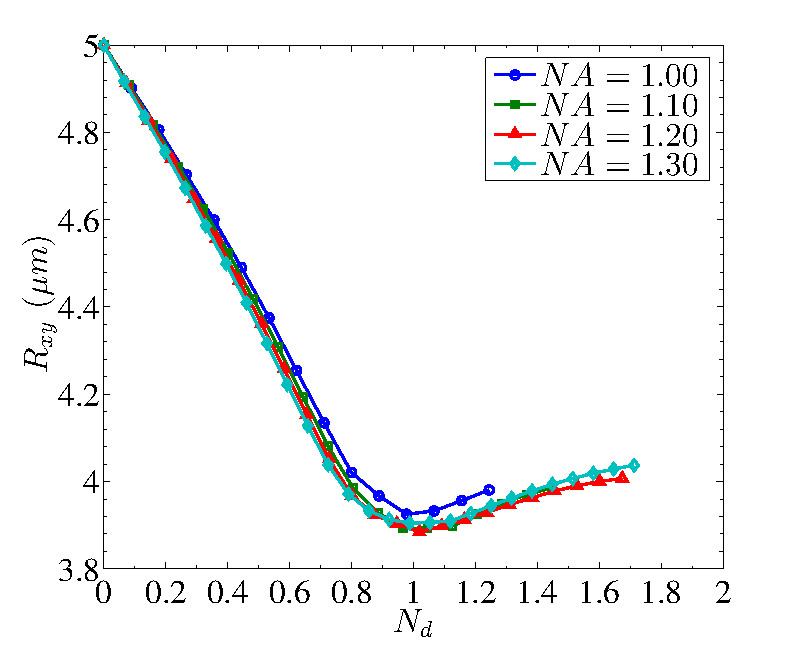}}\centering
\subfigure[][]{\includegraphics[width=0.45\textwidth, trim=0cm 0cm 1cm 0cm,clip=true]{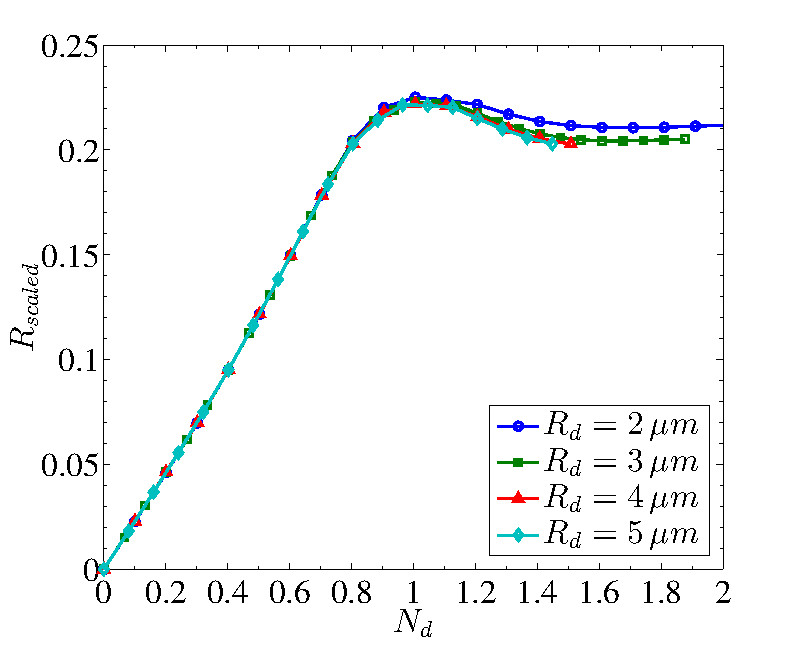}}\centering

\subfigure[][]{\includegraphics[width=0.45\textwidth, trim=0cm 0cm 1cm 0cm,clip=true]{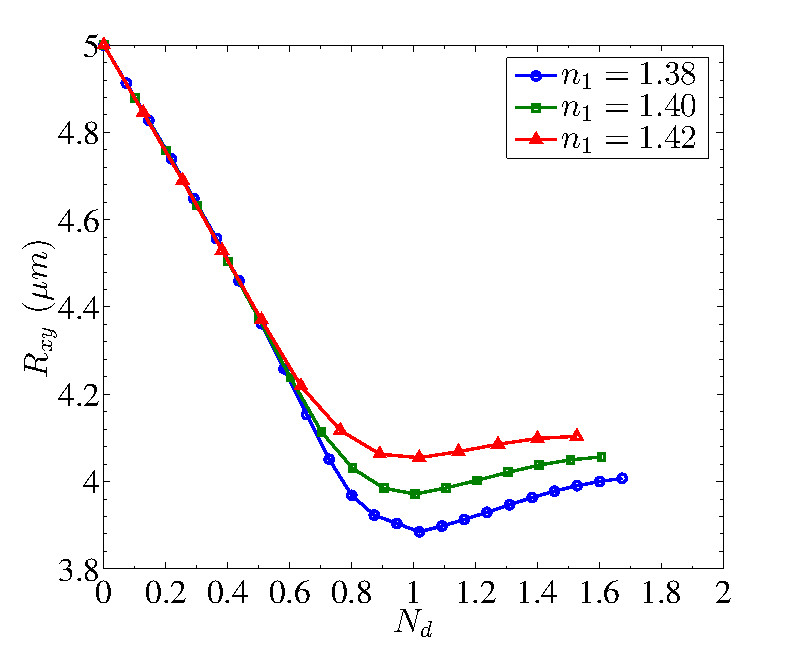}}\centering
\caption{Relationship between the deformation of a droplet and $N_{d}$ with varying (a) $NA$ for $R_{d}=5.0\,\mu m$ and $\gamma=10^{-6}\,Nm^{-1}$, (b) $R_{d}$ with $NA=1.10$ and $\gamma = 10^{-6}\,Nm^{-1}$ and (c) $n_{1}$ for a droplet with $R_{d}=5.0\,\mu m$ and $\gamma=10^{-6}\,Nm^{-1}$, and an optical trap with $NA = 1.20$.}
\label{fig:onelasergraphs_nd}
\end{figure}

The laser power at which a droplet undergoes a transition to a dumbbell-like shape, in a single optical trap, is found to be linearly dependent on both the initial radius of the droplet $\left(R_{d}\right)$ and the interfacial tension $\left(\gamma\right)$, and inversely proportional to the refractive index ratio $\tilde{n}$. The relationship between the laser power and the numerical aperture is slightly more complicated. We find that the laser power $P_{0} \propto \exp\left(NA\right)$. These relationships allow us to define a dimensionless number which characterises when a droplet will deform into a dumbbell-like shape, for a given initial radius, surface tension, refractive index ratio and numerical aperture:
\begin{equation}
N_{d} = \frac{P_{0} \tilde{n}}{\gamma R_{d} c \exp\left(NA\right)}
\end{equation}
When $N_{d} \gtrsim 1.0$ the droplet deforms into a dumbbell-like shape, whereas for $N_{d} \lesssim 1.0$ the droplet only elongates in the direction of the propagation of light. Figure \ref{fig:onelasergraphs_nd}a shows this for a constant initial spherical radius $R_{d} = 5.0\,\mu m$ whilst varying the numerical aperture for the optical tweezer. Figure \ref{fig:onelasergraphs_nd}b shows the variation of $R_{xy}$ vs.\ $N_{d}$ for different starting radii, whilst using a numerical aperture of $NA = 1.10$. To obtain a data collapse for the graphs in Figure \ref{fig:onelasergraphs_nd}b we rescale the observed $R_{xy}$ values as:
\begin{equation}
R_{scale} = \frac{R_{d}-R_{xy}}{R_{d}}
\end{equation}
Figure \ref{fig:onelasergraphs_nd}c also shows the same trend for a variation in the refractive index of the droplet $n_{1}$. As it can be seen, the minimum in Figures \ref{fig:onelasergraphs_nd}a \& c and maximum in Figure \ref{fig:onelasergraphs_nd}b, indicating the transition to a dumbbell-like shape, occurs at $N_{d} \approx 1.0$.

The transition to a dumbbell-like shape can be explained in terms of a balance between the forces due to the optical tweezer and the interfacial tension of the droplet. At values of $N_{d} \lesssim 0.7$ the interfacial tension energy is stronger than that of the optical tweezer, resulting in a linear deformation of the droplet. Between $N_{d} \approx 0.7$ and $1.0$ the forces exerted by the laser field begin to overcome the interfacial tension of the droplet, until $N_{d} \approx 1.0$ at which the optical forces dominate. The droplet interface then conforms locally to the iso-intensity contours of the laser beam, which results in the observed dumbbell-like geometry.

Beyond a certain laser power our model is not capable of accurately predicting the converged shapes of droplets. This breakdown of the model is due to a single radial ``spoke'' intersecting the interface multiple times. This critical point for our numerical model will occur as the droplet elongates further and the dumbbell-like shape becomes more defined. Our model as described here is implemented using a spherical coordinate system, where each ``spoke'' radiates out from the origin of the system. Refining the number of ``spokes'' can allow the model to approach this critical point more closely (albeit at larger computational cost), but the model as presently formulated is unable to pass this critical point.

\subsection{Multiple Optical Traps}
In addition to calculating the shape of a droplet in a single optical trap, our model is able to successfully calculate how a droplet will deform in multiple optical traps. To our knowledge this is the first time such a model has been presented. Each optical tweezer is moved  slowly outwards by $0.25\,\mu m$ in the appropriate directions, and we ensure that the droplet shape has converged at each stage before the lasers are moved again.

\begin{figure}[htbp]
\centering
\subfigure{\includegraphics[width=0.30\textwidth, trim=0cm 0cm 0cm 0cm,clip=true]{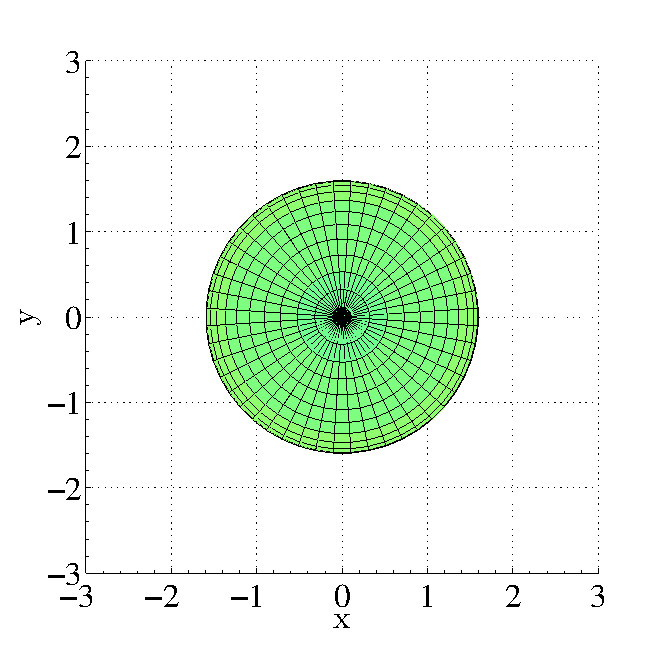}}\centering
\subfigure{\includegraphics[width=0.30\textwidth, trim=0cm 0cm 0cm 0cm,clip=true]{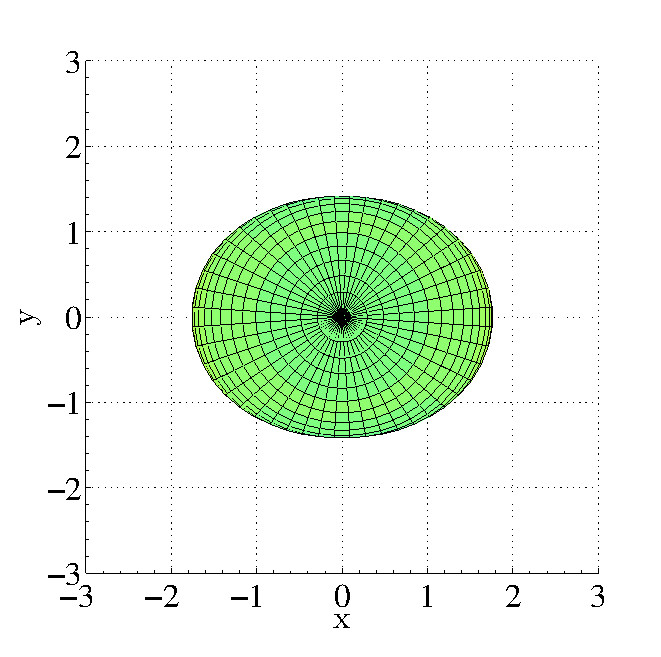}}\centering
\subfigure{\includegraphics[width=0.30\textwidth, trim=0cm 0cm 0cm 0cm,clip=true]{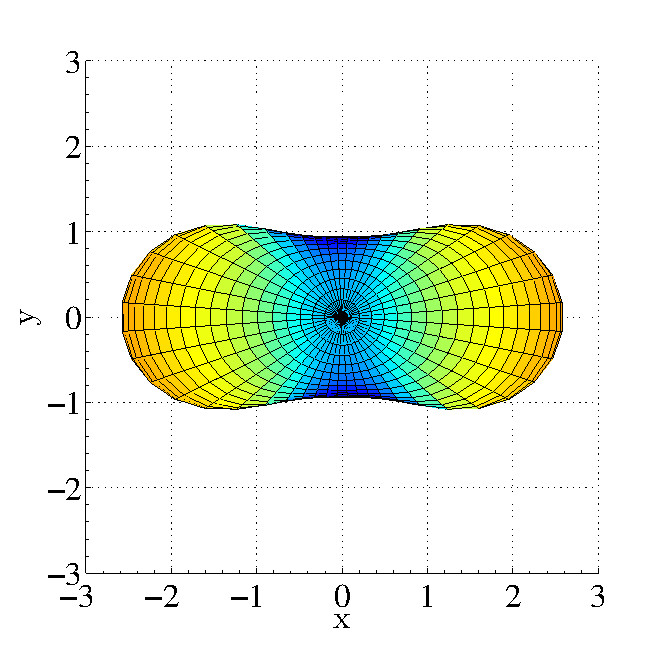}}\centering

\setcounter{subfigure}{0}
\subfigure[][]{\includegraphics[width=0.30\textwidth, trim=0cm 0cm 0cm 0cm,clip=true]{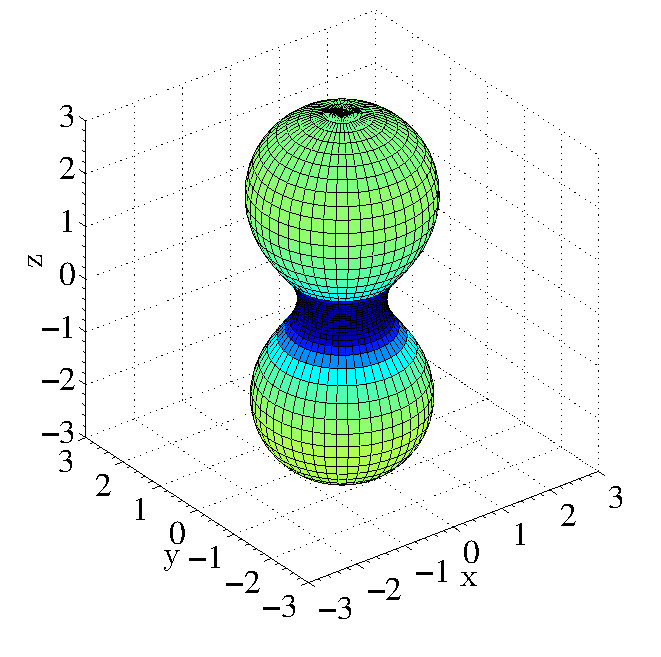}}\centering
\subfigure[][]{\includegraphics[width=0.30\textwidth, trim=0cm 0cm 0cm 0cm,clip=true]{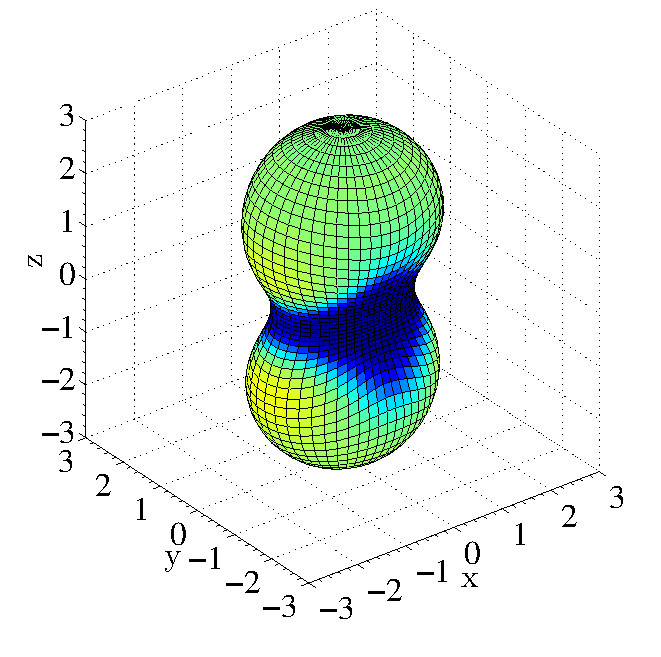}}\centering
\subfigure[][]{\includegraphics[width=0.30\textwidth, trim=0cm 0cm 0cm 0cm,clip=true]{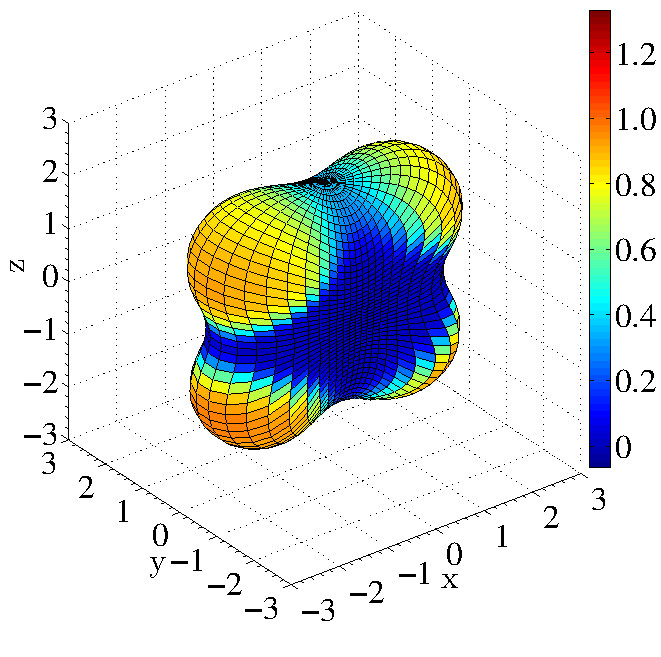}}\centering
\caption{Deformation of a $R_{d} = 2.0\,\mu m$ droplet using two optical traps each with $P_{0} = 0.06\,W$ and $NA=1.20$. The top images show the two-dimensional $xy$ projections and the bottom images show the corresponding three-dimensional geometries. Both lasers are positioned at the origin for (a). The positions of the lasers are then $\left(1.0,[0,\pi],0\right)\,\mu m$ \& $\left(2.0,[0,\pi],0\right)\,\mu m$ in polar coordinates for (b) and (c) respectively. The colour scheme represents the variation in mean curvature, in units of $\mu m^{-1}$.}
\label{fig:twolasers}
\end{figure}

\begin{figure}[htbp]
\centering
\subfigure{\includegraphics[width=0.30\textwidth, trim=0cm 0cm 0cm 0cm,clip=true]{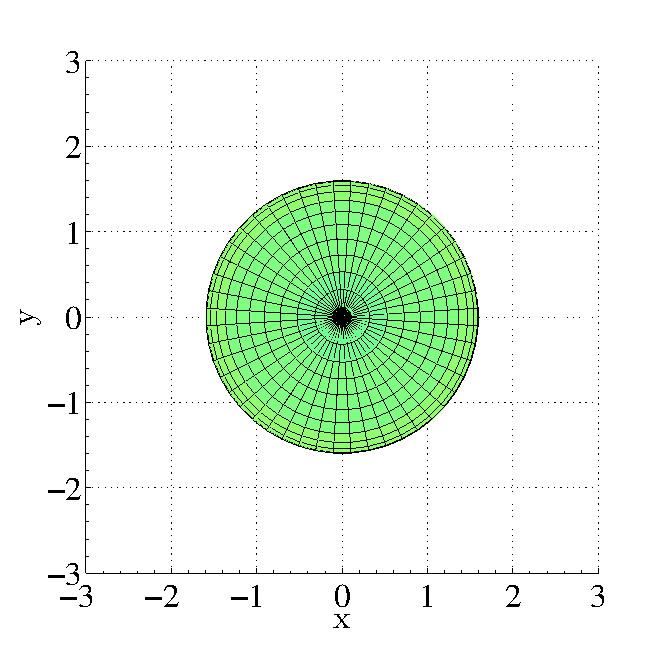}}\centering
\subfigure{\includegraphics[width=0.30\textwidth, trim=0cm 0cm 0cm 0cm,clip=true]{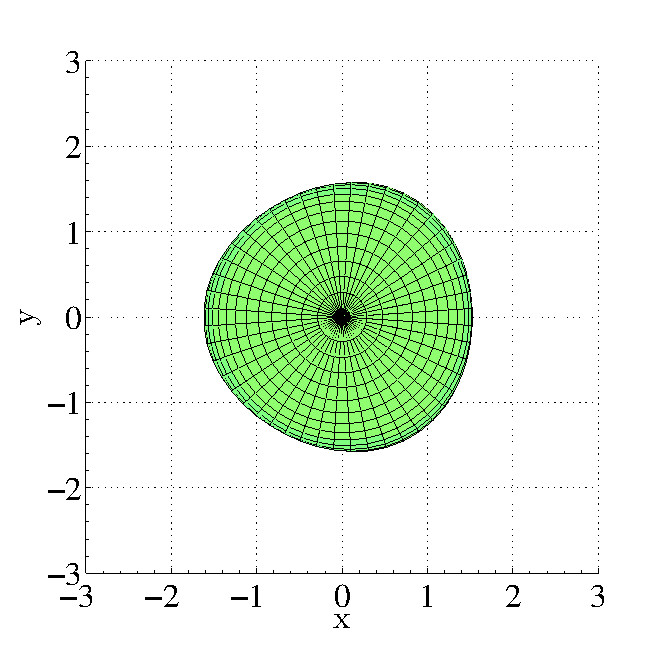}}\centering
\subfigure{\includegraphics[width=0.30\textwidth, trim=0cm 0cm 0cm 0cm,clip=true]{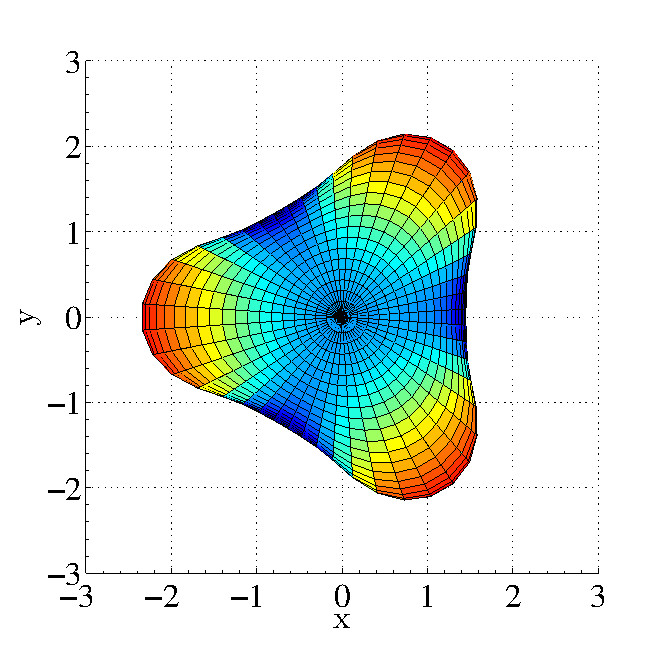}}\centering

\setcounter{subfigure}{0}
\subfigure[][]{\includegraphics[width=0.30\textwidth, trim=0cm 0cm 0cm 0cm,clip=true]{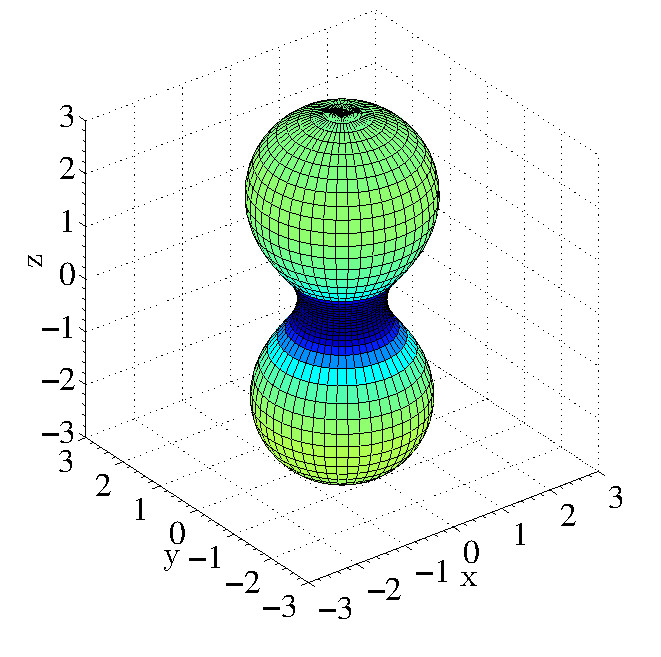}}\centering
\subfigure[][]{\includegraphics[width=0.30\textwidth, trim=0cm 0cm 0cm 0cm,clip=true]{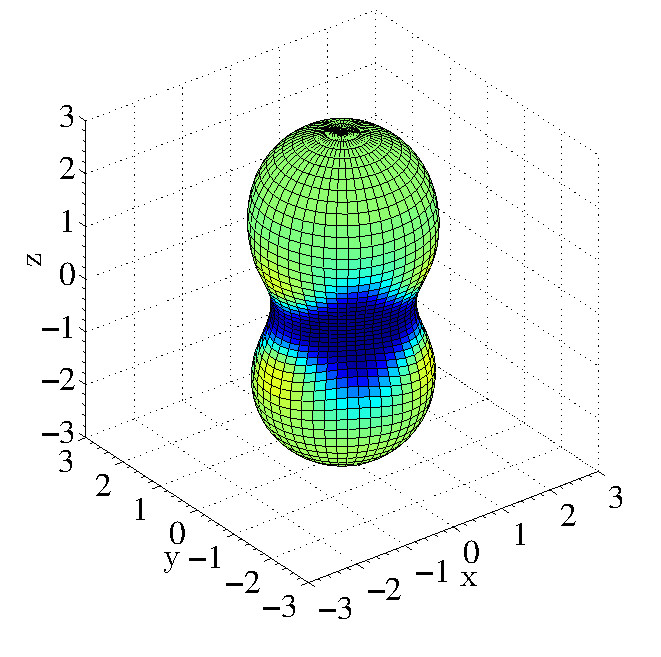}}\centering
\subfigure[][]{\includegraphics[width=0.30\textwidth, trim=0cm 0cm 0cm 0cm,clip=true]{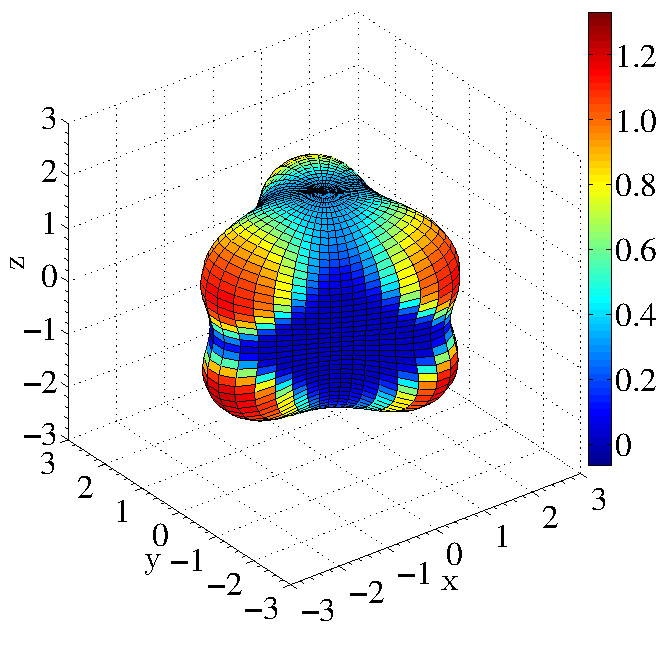}}\centering
\caption{Deformation of a $R_{d} = 2.0\,\mu m$ droplet using three optical traps each with $P_{0} = 0.04\, W$ and $NA=1.20$. The top images represent the two-dimensional $xy$ projections and the bottom images show the corresponding three-dimensional geometries. All lasers are positioned at the origin for (a). The positions of the lasers are then $\left(1.0,[\frac{\pi}{3},\pi,\frac{5\pi}{3}],0\right)\,\mu m$ \& $\left(2.0,[\frac{\pi}{3},\pi,\frac{5\pi}{3}],0\right)\,\mu m$ in polar coordinates for (b) and (c) respectively. The colour scheme represents the variation in mean curvature, in units of $\mu m^{-1}$.}
\label{fig:threelasers}
\end{figure}

\begin{figure}[htbp]
\centering
\subfigure{\includegraphics[width=0.30\textwidth, trim=0cm 0cm 0cm 0cm,clip=true]{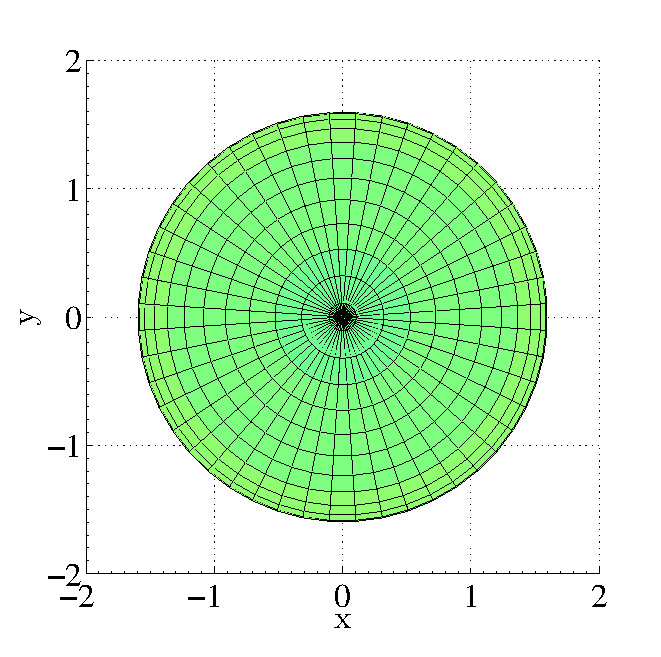}}\centering
\subfigure{\includegraphics[width=0.30\textwidth, trim=0cm 0cm 0cm 0cm,clip=true]{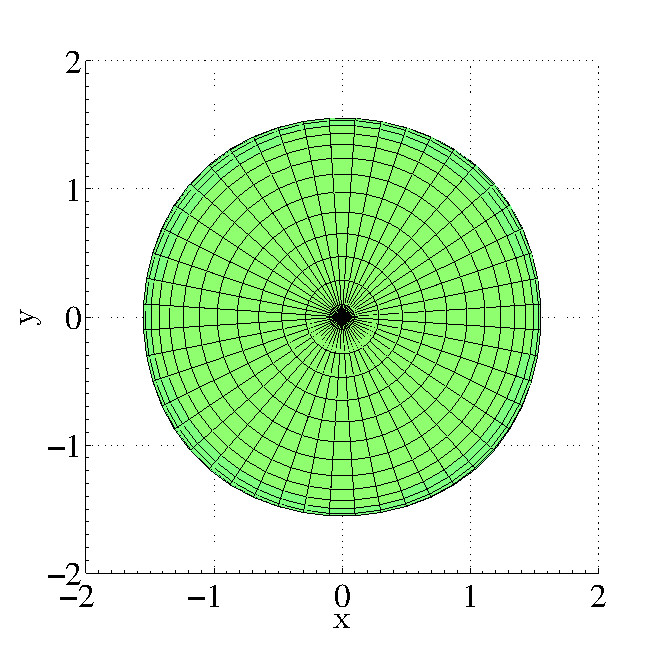}}\centering
\subfigure{\includegraphics[width=0.30\textwidth, trim=0cm 0cm 0cm 0cm,clip=true]{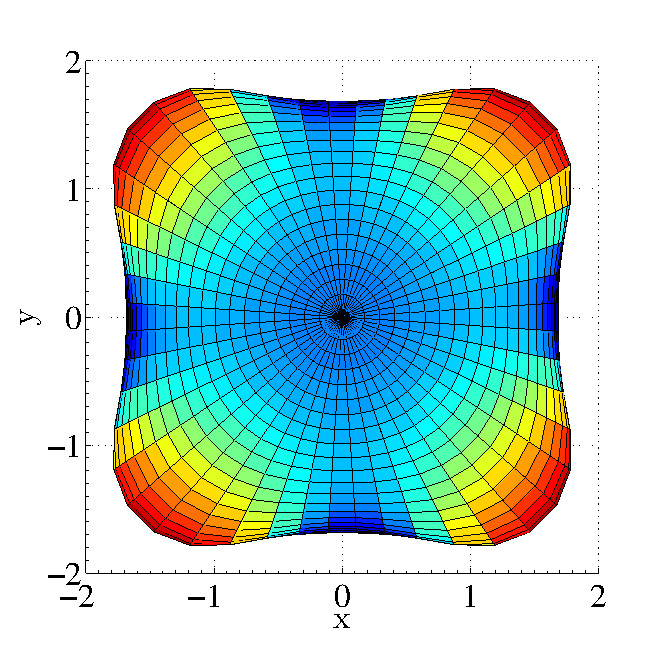}}\centering

\setcounter{subfigure}{0}
\subfigure[][]{\includegraphics[width=0.30\textwidth, trim=0cm 0cm 0cm 0cm,clip=true]{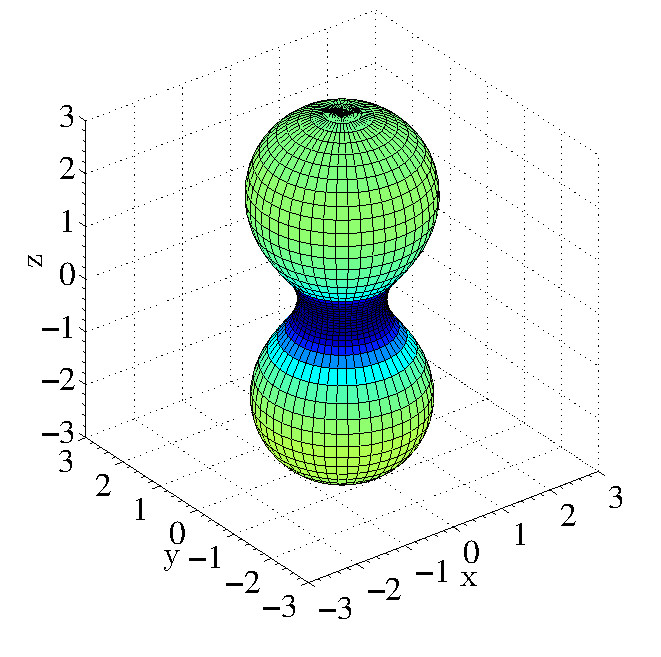}}\centering
\subfigure[][]{\includegraphics[width=0.30\textwidth, trim=0cm 0cm 0cm 0cm,clip=true]{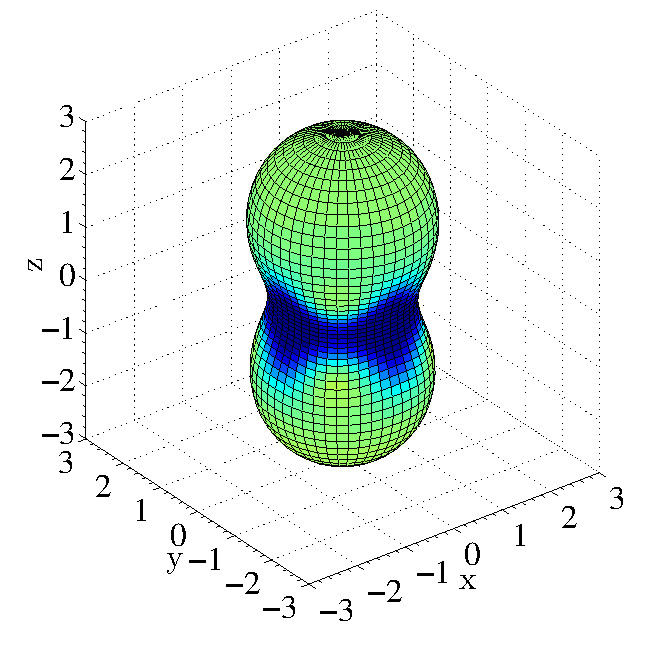}}\centering
\subfigure[][]{\includegraphics[width=0.30\textwidth, trim=0cm 0cm 0cm 0cm,clip=true]{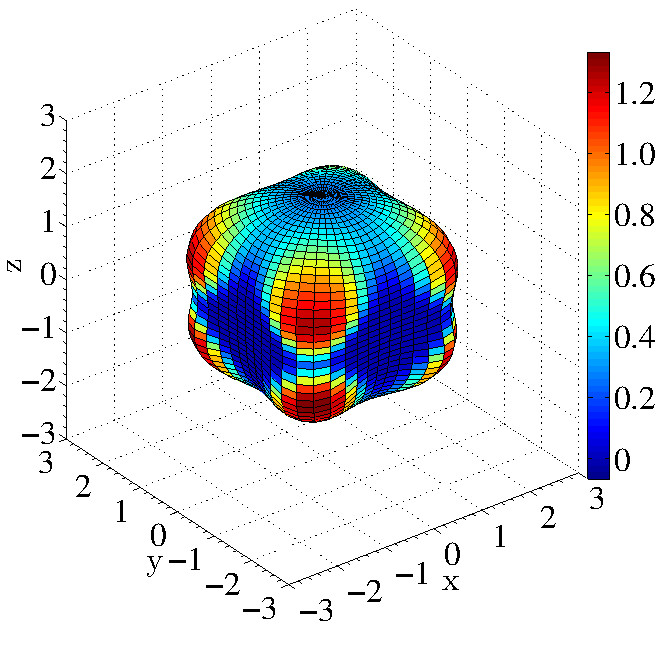}}\centering
\caption{Deformation of a $R_{d} = 2.0\,\mu m$ droplet using four optical traps each with $P_{0} = 0.03\, W$ and $NA=1.20$. The top images represent the two-dimensional $xy$ projections and the bottom images show the corresponding three-dimensional geometries. All lasers are positioned at the origin for (a). The positions of the lasers are then $\left(1.0,[\frac{\pi}{4},\frac{3\pi}{4},\frac{5\pi}{4},\frac{7\pi}{4}],0\right)\,\mu m$ \& $\left(2.0,[\frac{\pi}{4},\frac{3\pi}{4},\frac{5\pi}{4},\frac{7\pi}{4}],0\right)\,\mu m$ in polar coordinates for (b) and (c) respectively. The colour scheme represents the variation in mean curvature, in units of $\mu m^{-1}$.}
\label{fig:fourlasers}
\end{figure}

Figures \ref{fig:twolasers}, \ref{fig:threelasers} \& \ref{fig:fourlasers} show the converged shapes of an $R_{d} = 2.0\,\mu m$ droplet in two, three and four optical traps, respectively. In each Figure, moving from left to right, there is an increase in the separation between each optical trap. Initially, all traps are positioned at the centre of the droplet. The total laser power acting on the droplet is kept constant at $P_{total}=0.12\,W$ and is equally shared between the total number of lasers being modelled. The interfacial tension and numerical aperture were also kept constant at $\gamma = 10^{-6}\,Nm^{-1}$ and $NA = 1.20$ respectively. The top row of each figure shows the two-dimensional $xy$ projections of the deformed droplet, as might be seen in a brightfield microscope image. The chosen colour scheme indicates the variation in the calculated mean curvature; as the separation of each laser increases there is a larger variation in the mean curvature. For the largest separation between optical traps,
 presented in Figures \ref{fig:twolasers}c, \ref{fig:threelasers}c \& \ref{fig:fourlasers}c the mean curvature is highest at the ``tips'' or ``corners'' of the shapes formed, where the tightly focused lasers are positioned.

As mentioned above, a major benefit of these numerical results is that it is possible to view the three-dimensional geometries of the deformed droplet. These are shown in the bottom row of each of the Figures \ref{fig:twolasers} - \ref{fig:fourlasers}. To date such observations have not been achieved experimentally. When all lasers are positioned at the centre of the droplet then $P_{total}=P_{0}$ and we obtain a value of $N_{d}=2.18$ for the given $R_{d}$,  $\gamma$ and $NA$. Hence this is beyond the transition to a dumbbell-like shape, as shown in Figures \ref{fig:twolasers} - \ref{fig:fourlasers}a. The hour-glass connecting the two spherical caps has a negative mean curvature. As each individual laser is then moved outwards from the centre, the surface of the droplet at the ``tips'' or ``corners'' retains this concave shape. This observation is interesting since from just viewing the two-dimensional projections one might assume the surfaces to be convex along the axis of propagation of light.

It is worth mentioning that for suitably selected parameters we can observe a range of deformed droplets with a high proportion of negative mean curvature. For example, a droplet with initial radius $R_{d} = 5.0\mu m$ and interfacial tension $\gamma = 10^{-6}\,Nm^{-1}$, deformed in four optical traps each with a laser power of $P_{0}=0.1\,W$ and numerical aperture $NA=0.8$ has negative mean curvature on the top and bottom faces, in addition to that seen on the side faces in Figure \ref{fig:fourlasers}. This indicates that the deformation of droplets by optical forces is extremely sensitive to the selected parameters.

\begin{figure}[htbp]
\centering
\subfigure{\includegraphics[width=0.25\textwidth, trim=0cm 0cm 0cm 0cm,clip=true]{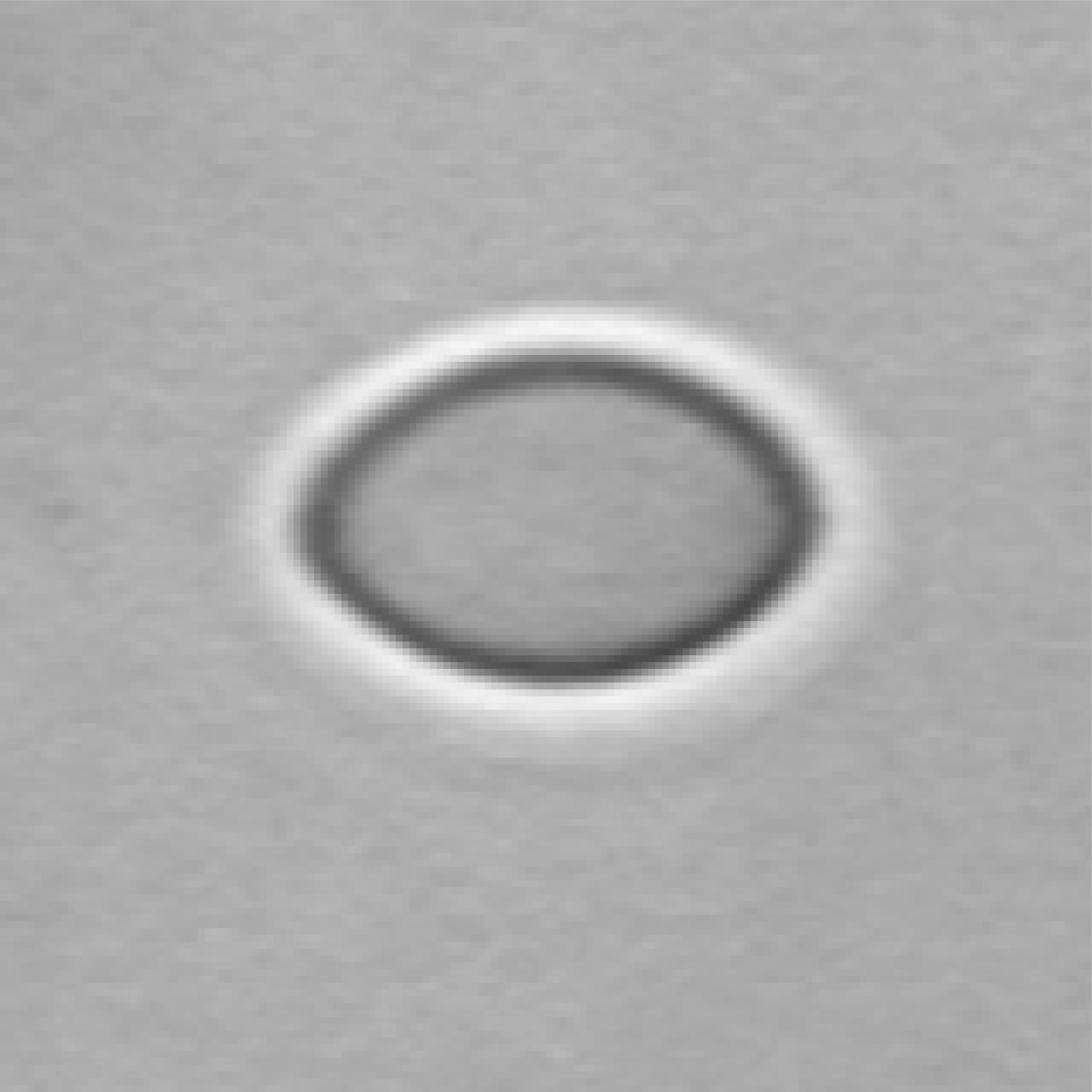}}\centering \hspace{0.6cm}
\subfigure{\includegraphics[width=0.25\textwidth, trim=0cm 0cm 0cm 0cm,clip=true]{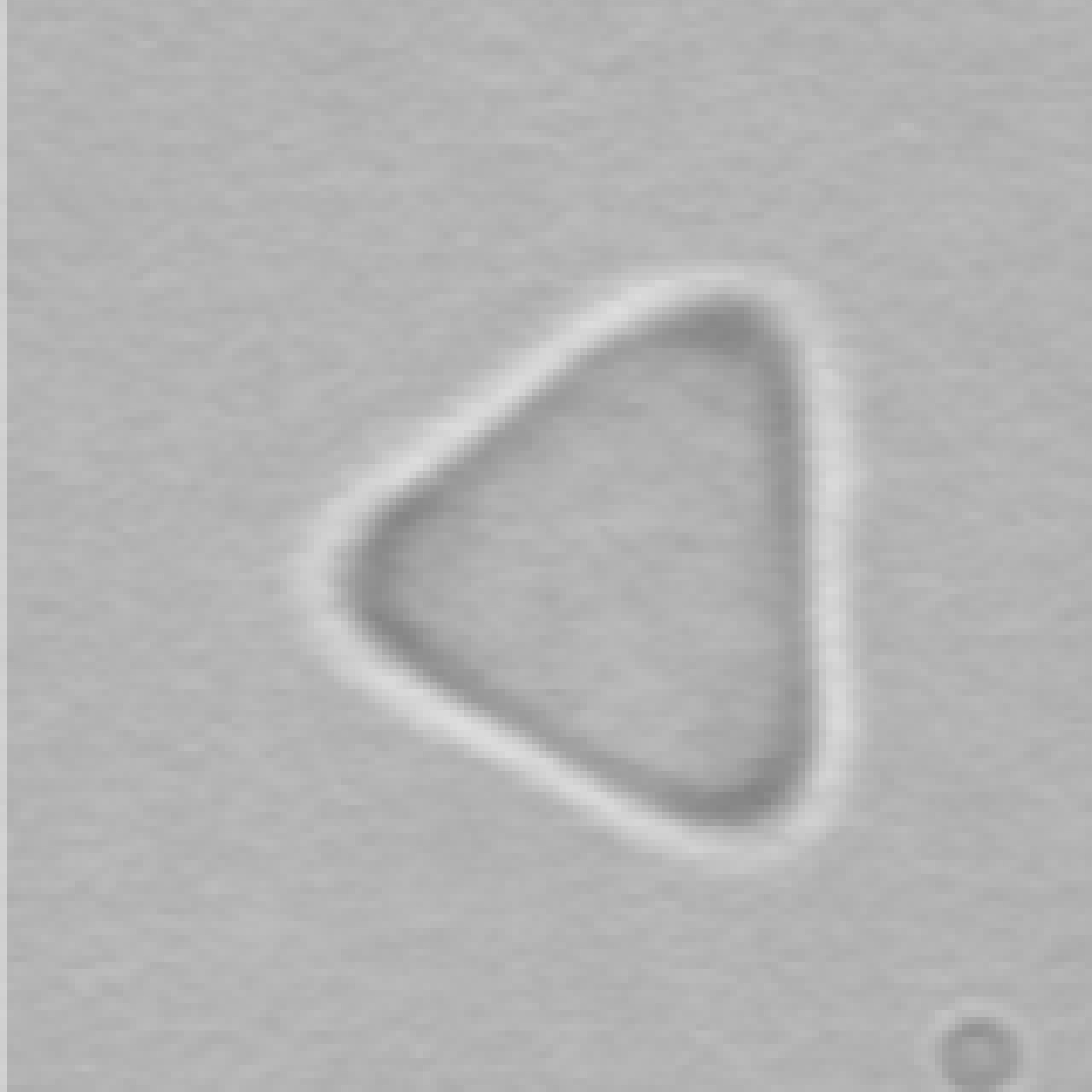}}\centering \hspace{0.6cm}
\subfigure{\includegraphics[width=0.25\textwidth, trim=0cm 0cm 0cm 0cm,clip=true]{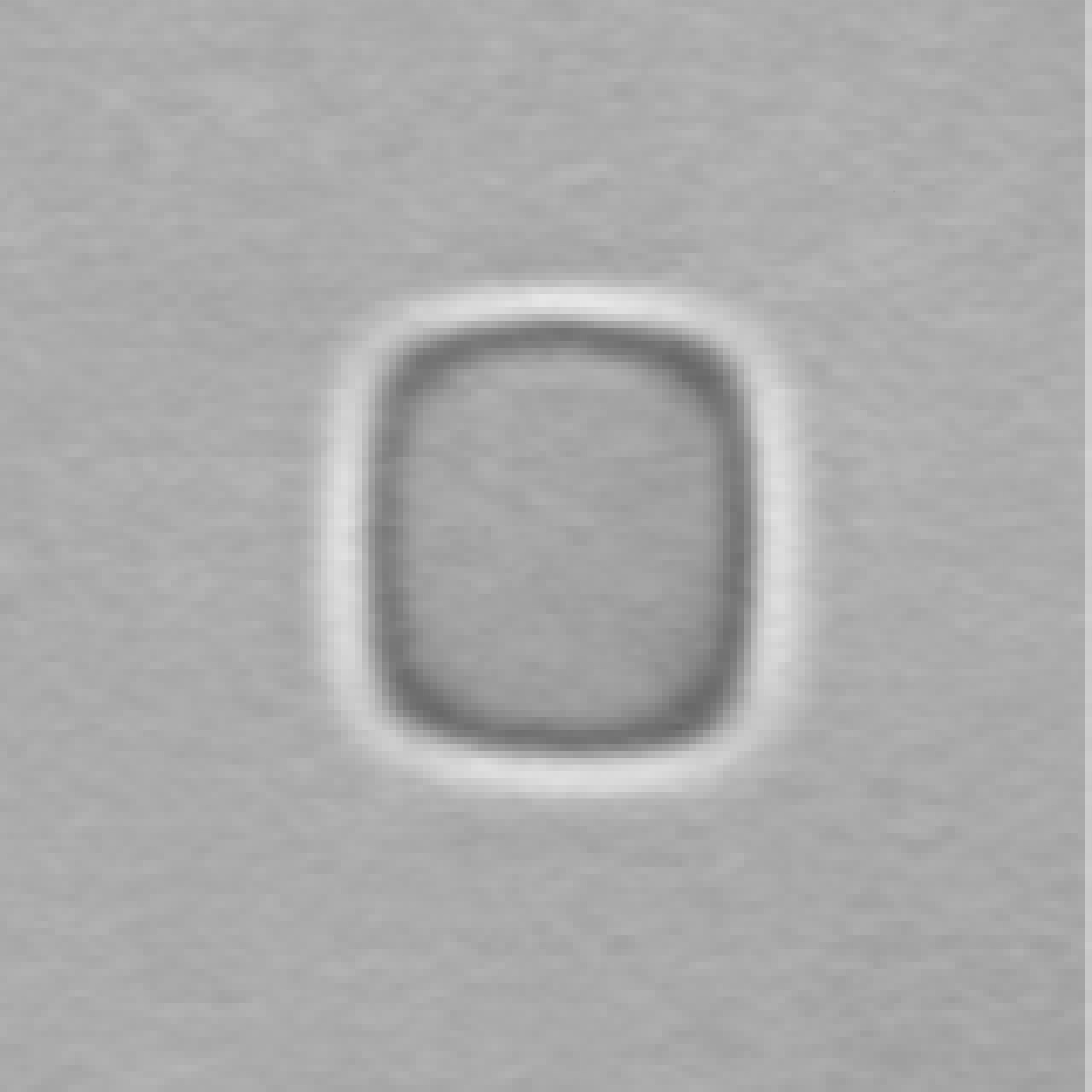}}\centering

\subfigure{\includegraphics[width=0.30\textwidth, trim=0cm 0cm 0cm 0cm,clip=true]{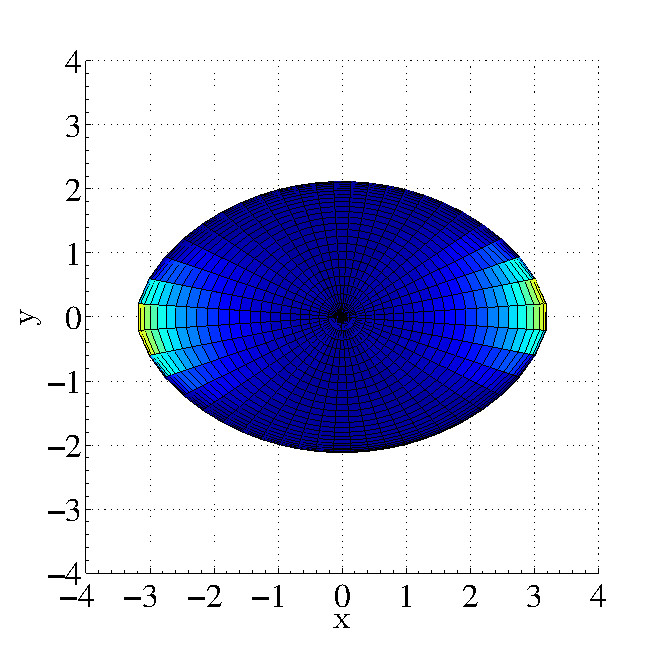}}\centering
\subfigure{\includegraphics[width=0.30\textwidth, trim=0cm 0cm 0cm 0cm,clip=true]{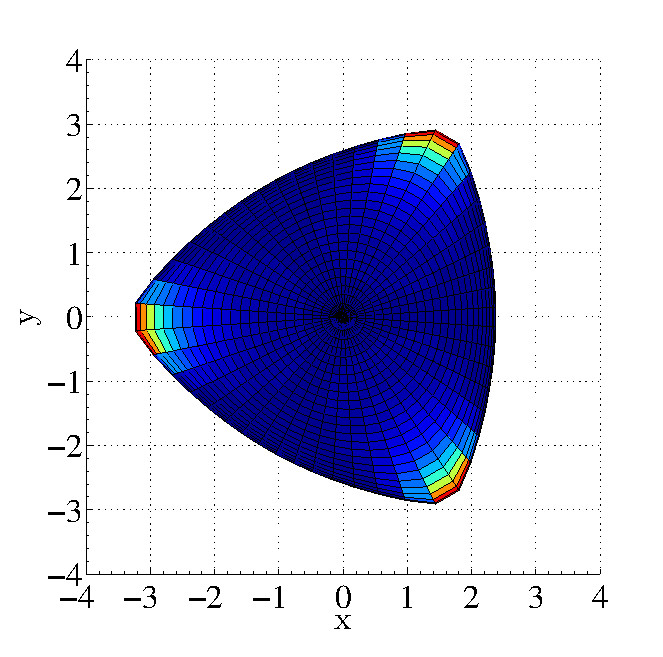}}\centering
\subfigure{\includegraphics[width=0.30\textwidth, trim=0cm 0cm 0cm 0cm,clip=true]{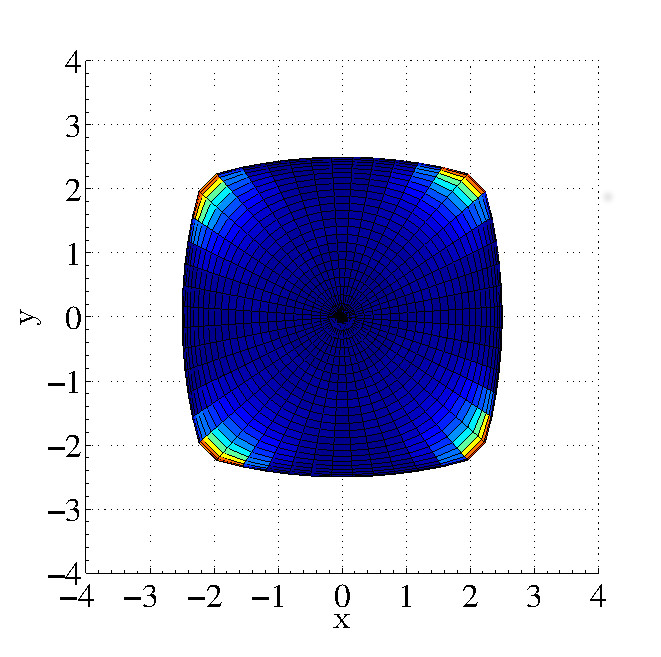}}\centering

\setcounter{subfigure}{0}
\subfigure[][]{\includegraphics[width=0.30\textwidth, trim=0cm 0cm 0cm 0cm,clip=true]{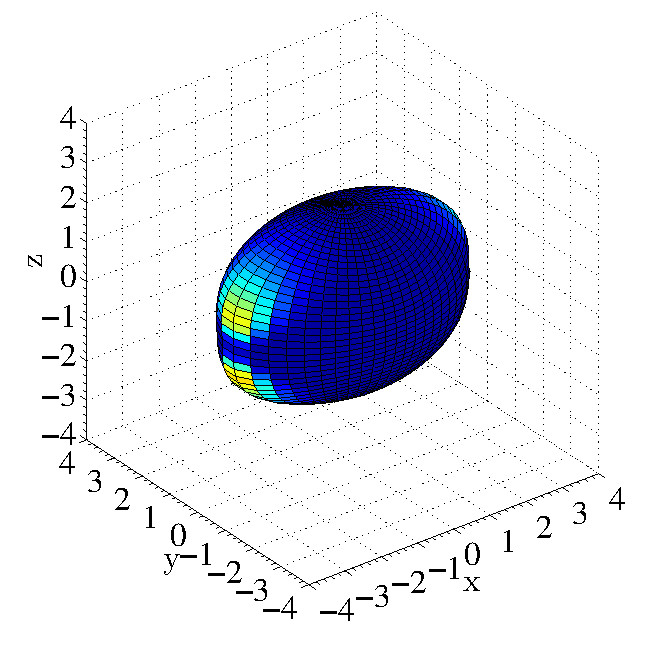}}
\subfigure[][]{\includegraphics[width=0.30\textwidth, trim=0cm 0cm 0cm 0cm,clip=true]{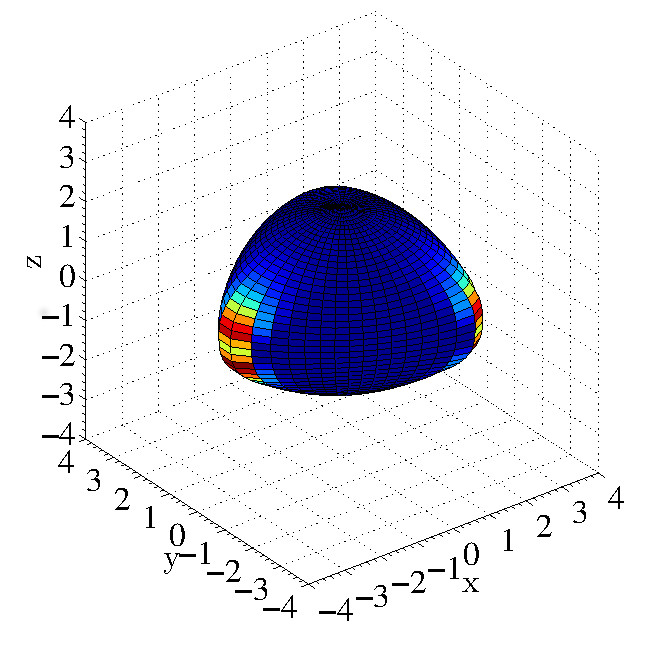}}
\subfigure[][]{\includegraphics[width=0.30\textwidth, trim=0cm 0cm 0cm 0cm,clip=true]{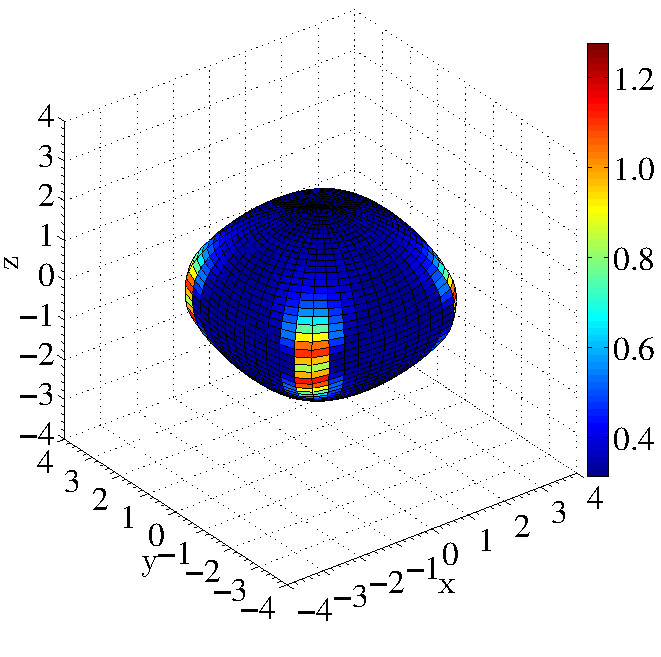}}
\caption{Deformation of a $R_{d} = 2.5\,\mu m$ droplet with $\gamma = 10^{-6}\,Nm^{-1}$ using two, three and four optical traps in (a)-(c) respectively. The total combined optical power is $P_{total}=24\,mW$ with a numerical aperture of $NA=1.20$ for each laser. The focus of each laser is a lateral distance $3\,\mu m$ from the centre of symmetry of the experiment. The top row represents the two-dimensional $xy$ projections observed experimentally (Ref.~\cite{Bain:06} -- Reproduced by permission of The Royal Society of Chemistry \texttt{http://pubs.rsc.org/en/Content/ArticleLanding/2006/CC/b610060k}), the middle row are our predicted shapes and the bottom row shows the corresponding three-dimensional geometries. The colour scheme for our calculated structures represents the variation in mean curvature, in units of $\mu m^{-1}$.}
\label{fig:expvalues}
\end{figure}

As an experimental validation of our model, we compare our calculated two-dimensional shapes of a deformed emulsion droplet to those presented in the work of Ward \emph{et al.}~\cite{Bain:06}. To summarise the parameters used in their experiments: an approximate value of $R_{d}=2.5\,\mu m$ was taken for the initial spherical geometry of the droplet, and an interfacial tension of $\gamma \approx 10^{-6}\,Nm^{-1}$ was reported. For the optical traps, each laser had a numerical aperture of $NA=1.20$ and a total combined laser power of $P_{total}=24\,mW$ was equally distributed between the total number of lasers. The experimentally observed shapes and steady-state shapes predicted by our model are presented in the first and second rows of Figure \ref{fig:expvalues} respectively, for an increasing number of optical traps.

As mentioned above, the highest calculated mean curvature values occur near the focal positions of lasers. It can be seen from Figure \ref{fig:expvalues} that using the experimental parameters of Ward~\emph{et al.} we obtain droplets with conical ends. The high electric field at the laser foci exerts a very strong force on the interface, and when combined with volume conservation and surface curvature considerations, this results in a locally very small radius of curvature at the tips of the shape, as previously modelled by Stone~\emph{et al.}~\cite{Stone:99} based on observations of ``Taylor cones'' in static electric fields~\cite{Taylor1964}.

The bottom row of Figure \ref{fig:expvalues} show the three-dimensional geometries of these emulsion droplets. Unlike the structures presented in Figures \ref{fig:twolasers}-\ref{fig:fourlasers}, the ``tips'' or ``corners'' of each droplet remain convex, as expected from the value of $N_{d} = 0.35$ deduced from the parameters reported by Ward \emph{et al.}~\cite{Bain:06}. Hence when all the lasers are positioned at the centre of the droplet at the start of the experiment, the droplet remains approximately spherical, and does not take on the dumbbell-like structure discussed earlier. This in turn reduces the deformation of the droplet along the light propagation axis at the ``tips''/``corners'' as each laser is moved outwards.

There is a very strong agreement between our theoretical predictions of the two-dimensional $xy$ projections of the deformed droplets and those obtained experimentally. We hope that this work will prompt experimental studies to visualise the three-dimensional configurations as a function of the physical parameters explored in this paper.

\section{Conclusion}

In conclusion we have developed a theoretical framework to compute three-dimensional equilibrium shapes of liquid droplets with ultra-low interfacial tension in one or more optical traps. Taking a cue from experiments, we assume an isotropic, temperature-independent interfacial tension coefficient $\gamma$. The optical traps were described using a far-field scalar model of a tightly focused Gaussian beam, within the Rayleigh-Gans regime.

The equilibrium droplet shape arises as a result of the interfacial tension trying to minimise the droplet surface area, the internal fluid pressure resisting such a change, and the external optical pressure causing local deformations, without any volumetric change. Using this model we numerically compute the droplet shapes as a function of both droplet and laser parameters e.g.\ interfacial tension, initial droplet size, laser power and numerical aperture.

We obtain droplet shapes similar to those obtained experimentally by Ward \emph{et al.} for similar parameter values. The close agreement between theoretical predictions of two-dimensional projections in the $xy$ plane and experiments for known geometries and parameter values gives us confidence in the correctness of our predicted three-dimensional droplet shapes. It is worth noting that these predictions of three-dimensional droplet shapes for large deformations (where the linear response breaks down) have not been reported in the literature. Experiments are currently under development to measure three-dimensional shapes for comparison with our model

Further, the theoretical work revealed a few surprises along the way. For example, in a single optical trap as the laser power is increased the trapped droplet not only elongates in the direction of light propagation, but also deforms into a dumbbell-like shape. We have predicted the onset of this transformation in terms of a dimensionless deformation number $N_{d}$, obtained using dimensional arguments balancing antagonistic surface tension and optical forces. The mathematical expression has been substantiated using a data collapse of the numerical solution of the shape equations.

We have also predicted droplet shapes having a total negative mean curvature along its faces for optical traps having a high intensity. Though for some configurations (e.g.\ four laser traps) this may be intuitive, the existence of droplet shapes having a total negative mean curvature in a single trap in not immediately obvious.

\section*{Acknowledgements}

The authors wish to acknowledge funding from EPSRC via grant EP/I013377/1. We thank Oscar Ces and Andrew Ward, along with all others in the optonanofluidics group for their helpful discussions and suggestions. The authors acknowledge helpful discussions with P. D. Olmsted, and others of the external advisory board of the optonanofluidics project. BC thanks the Newton Institute for support.


\end{document}